\documentclass[journal]{IEEEtran} 
\usepackage{xcolor,soul,framed} 
\colorlet{shadecolor}{yellow}
\usepackage[pdftex]{graphicx}
\graphicspath{{../pdf/}{../jpeg/}}
\DeclareGraphicsExtensions{.pdf,.jpeg,.png}
\usepackage[cmex10]{amsmath}
\usepackage{csquotes}
\usepackage[T1]{fontenc} 
\usepackage{textcomp} 
\usepackage{array}
\usepackage{mdwmath}
\usepackage{mdwtab}
\usepackage{eqparbox}
\usepackage{url}
\usepackage{hyperref}
\usepackage{subcaption}
\usepackage{booktabs}
\usepackage{pifont}
\usepackage{verbatim}
\usepackage{float}
\usepackage{hyperref}
\usepackage{xcolor}
\usepackage{soul}
\usepackage{booktabs} 
\usepackage{caption} 
\usepackage{bm}
\usepackage{amsmath} 
\usepackage{amssymb}
\usepackage{siunitx}

\usepackage{graphicx}
\usepackage{subcaption}
\usepackage{cite}
\ifCLASSINFOpdf
\else
\fi
\begin{document}
%
\title{Amyloid Oligomers as Obstacle for Glutamate Diffusion in Synaptic Cleft}
%
%
%

     \title{A Molecular Communication Perspective of Alzheimer's Disease: Impact of Amyloid Beta Oligomers on Glutamate Diffusion in the Synaptic Cleft}
  \author{Nayereh FallahBagheri,~\IEEEmembership{Student Member,~IEEE} and
      \"Ozg\"ur B.~Akan,~\IEEEmembership{Fellow,~IEEE}\\

  \thanks{The authors are with the Center for neXt-generation Communications (CXC), Department of Electrical and Electronics Engineering, Ko\c{c} University, Istanbul 34450, Turkey  (e-mail: \{nbagheri23, akan\}@ku.edu.tr).
  
  O. B. Akan is also with the Internet of Everything (IoE) Group, Electrical Engineering Division, Department of Engineering, University of Cambridge, Cambridge CB3 0FA, UK (email: oba21@cam.ac.uk).
  
    This work was supported in part by the AXA Research Fund (AXA Chair
    for Internet of Everything at Ko\c{c} University)}
 }

\maketitle

\begin{abstract}

Molecular communication (MC) within the synaptic cleft is vital for neurotransmitter diffusion, a process critical to cognitive functions. In Alzheimer’s Disease (AD), beta-amyloid oligomers (A$\beta$os) disrupt this communication, leading to synaptic dysfunction. This paper investigates the molecular interactions between glutamate, a key neurotransmitter, and A$\beta$os within the synaptic cleft, aiming to elucidate the underlying mechanisms of this disruption. Through stochastic modeling, we simulate the dynamics of A$\beta$os and their impact on glutamate diffusion. The findings, validated by comparing simulated results with existing experimental data, demonstrate that A$\beta$os serve as physical obstacles, hindering glutamate movement and increasing collision frequency. This impairment of synaptic transmission and long-term potentiation (LTP) by binding to receptors on the postsynaptic membrane is further validated against known molecular interaction behaviors observed in similar neurodegenerative contexts. The study also explores potential therapeutic strategies to mitigate these disruptions. By enhancing our understanding of these molecular interactions, this research contributes to the development of more effective treatments for AD, with the ultimate goal of alleviating synaptic impairments associated with the disease.

\end{abstract}


%
\IEEEpeerreviewmaketitle

\section{Introduction}
\IEEEPARstart{A}{lzheimer}'s disease (AD) is a prevalent neurodegenerative disorder affecting over 32 million individuals globally \cite{diogo2022early}. This condition is marked by memory loss, cognitive decline, behavioral changes, and social challenges \cite{mebane20092009}. A hallmark of AD pathology is the presence of beta-amyloid oligomers (A$\beta$os), which trigger various toxic pathways leading to neuronal damage and widespread neural inflammation. These oligomers negatively impact synaptic functions and memory processes, contribute to abnormal tau protein modifications, and activate microglial cells \cite{tolar2021neurotoxic}.

The dysregulation of glutamatergic neural circuits is believed to play a critical role in the neurobiological foundation of AD \cite{matthews2021riluzole}. AD progression is notably characterized by the degeneration of large glutamatergic pyramidal neurons \cite{morrison2002selective}, \cite{hof1990quantitative}. A$\beta$os disrupt glutamate transporters \cite{li2009soluble}, alter activity-dependent glutamate release \cite{kamenetz2003app}, and decrease the surface expression of synaptic N-Methyl-D-Aspartate receptors (NMDARs) \cite{snyder2005regulation}, which are essential for physiological neurotransmission \cite{matthews2021riluzole}.

Glutamate, the brain's primary excitatory neurotransmitter, plays a vital role in cognitive functions such as learning and memory \cite{mcentee1993glutamate}. Its concentration in the brain varies between 5 and 15 millimoles per kilogram of brain tissue, depending on the region, making it the most abundant amino acid in the brain \cite{schousboe1981transport}.

Neurodegenerative and psychiatric disorders, including AD, are often linked to neuronal death caused by glutamate excitotoxicity or an imbalance between excitatory and inhibitory neuronal activities. This imbalance, particularly the abnormal ratio of excitatory/inhibitory inputs, is suggested to be a fundamental factor in neuropsychiatric conditions such as autism \cite{rubenstein2003model}, obsessive-compulsive disorder \cite{wu2012role}, and schizophrenia \cite{gao2015common}.

The dynamic and heterogeneous nature of A$\beta$os complicates therapeutic targeting, as highlighted by \cite{huang2020toxicity}. Despite extensive research, the transition from non-fibrillar to fibrillar oligomers remains poorly understood, indicating a significant gap in our knowledge of the biochemical mechanisms involved. Additionally, while various neurotoxic mechanisms of A$\beta$os have been proposed, their exact contributions to AD pathology are still unclear, necessitating further investigation.

Current studies emphasize the pivotal role of A$\beta$os in AD pathogenesis, particularly their strong correlation with neuronal loss and synaptic dysfunction, even more so than amyloid plaques. This understanding is essential for exploring how A$\beta$os disrupt glutamate diffusion and synaptic communication. However, clinical detection and quantification of A$\beta$os are challenging due to the lack of standardized, high-sensitivity assays, leading to inconsistent results across studies. This variability may be attributed to the heterogeneity of A$\beta$os and the diverse assay systems used. Moreover, the precise chemical composition and size-dependent properties of A$\beta$os, likely critical to their role in AD, require further clarification \cite{mroczko2018amyloid}.

Research indicates that A$\beta$os specifically bind to NMDARs containing GluN2B subunits and metabotropic glutamate receptors, causing synaptic disruptions. However, the exact receptors mediating the neurotoxicity of soluble A$\beta$os are not well-defined, posing challenges for developing targeted therapeutic strategies for AD. Furthermore, it remains unclear whether A$\beta$ protofibrils interact with specific NMDARs and mGluR1 or if A$\beta$ monomers engage with these receptors at all, highlighting another critical area for future research \cite{taniguchi2022amyloid}.

This study builds on these findings by employing a stochastic differential equation framework to model A$\beta$o-glutamate interactions. By incorporating variable oligomer sizes and simulating their impact on glutamate diffusion within a realistic synaptic environment, we aim to address these limitations and provide new insights into the mechanisms underlying synaptic dysfunction in AD.

The rest of this paper is structured as follows: In Sec. \ref{section:II}, the synaptic cleft and MC are introduced. The system is mathematically modeled in Sec. \ref{section:III}. In Sec. \ref{section:IV}, the interactions between A$\beta$os and glutamate, along with their impacts, are discussed. The final results are detailed in Sec. \ref{section:V}, and finally future directions and conclusions are outlined in Sec. \ref{section:VI}.


\section{The Synaptic Cleft and Molecular Communication}
\label{section:II}

\begin{figure}[ht]
\centering
\includegraphics[width=0.55\textwidth]{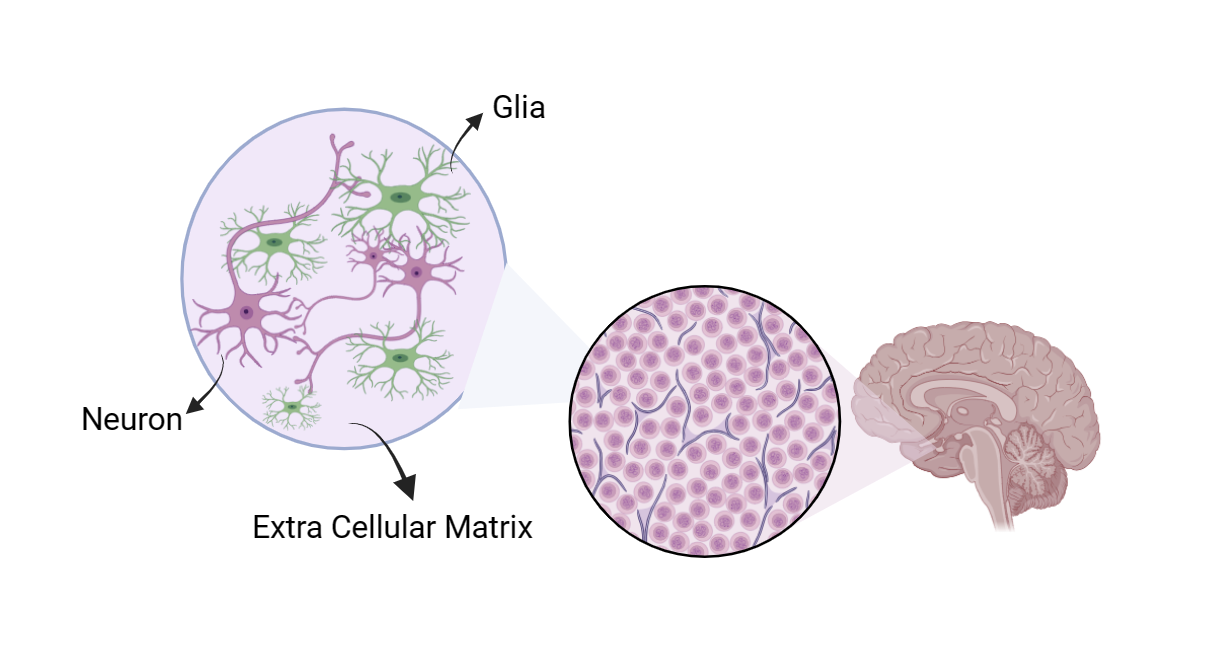}
\caption{Schematic illustration of the brain highlighting the distribution of glial cells, neurons, and the extracellular matrix.}
\label{fig:brain}
\end{figure}

The brain's myriad cells fall into two primary categories: information-processing neurons and the supportive glial cells, as illustrated in Fig. \ref{fig:brain}. Both are abundant and intricately designed, tightly arranged with tiny spaces between them. These intercellular gaps together form the brain's extracellular matrix (ECM), taking up roughly one-fifth of the brain's total volume \cite{owen2019effects}.\\

\begin{figure}[ht]
\centering
\includegraphics[width=0.4\textwidth]{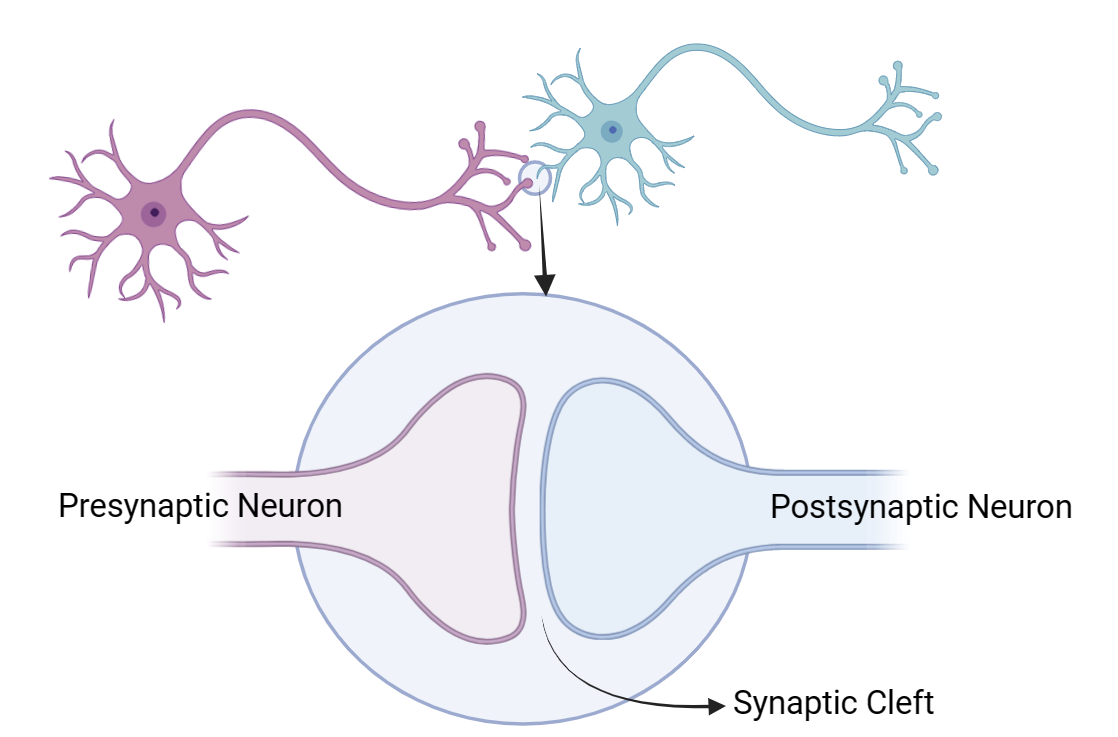}
\caption{Extracellular matrix between synapses (synaptic cleft)}
\label{fig:synaptic cleft}
\end{figure}

Neuronal synapses are critical junctions between neurons, facilitating the transmission of signals. These synapses consist of presynaptic and postsynaptic terminals, with the synaptic cleft forming the extracellular space between these adjacent cell membranes \cite{lucˇic2005morphological}, as illustrated in Fig. \ref{fig:synaptic cleft}.

\subsection{Structural Overview of the Synaptic Cleft}

The synaptic cleft is an essential component of neuronal communication, enabling the diffusion of neurotransmitters such as glutamate \cite{meldrum2000glutamate}. This region is heterogeneous \cite{cijsouw2018mapping}, with dimensions typically ranging from 0.3 to 0.5 $\mu$m in width and 1 to 1.5 $\mu$m in height \cite{zheng2017nanoscale}.

\begin{figure*}[t]
\centering
\includegraphics[width=0.7\textwidth]{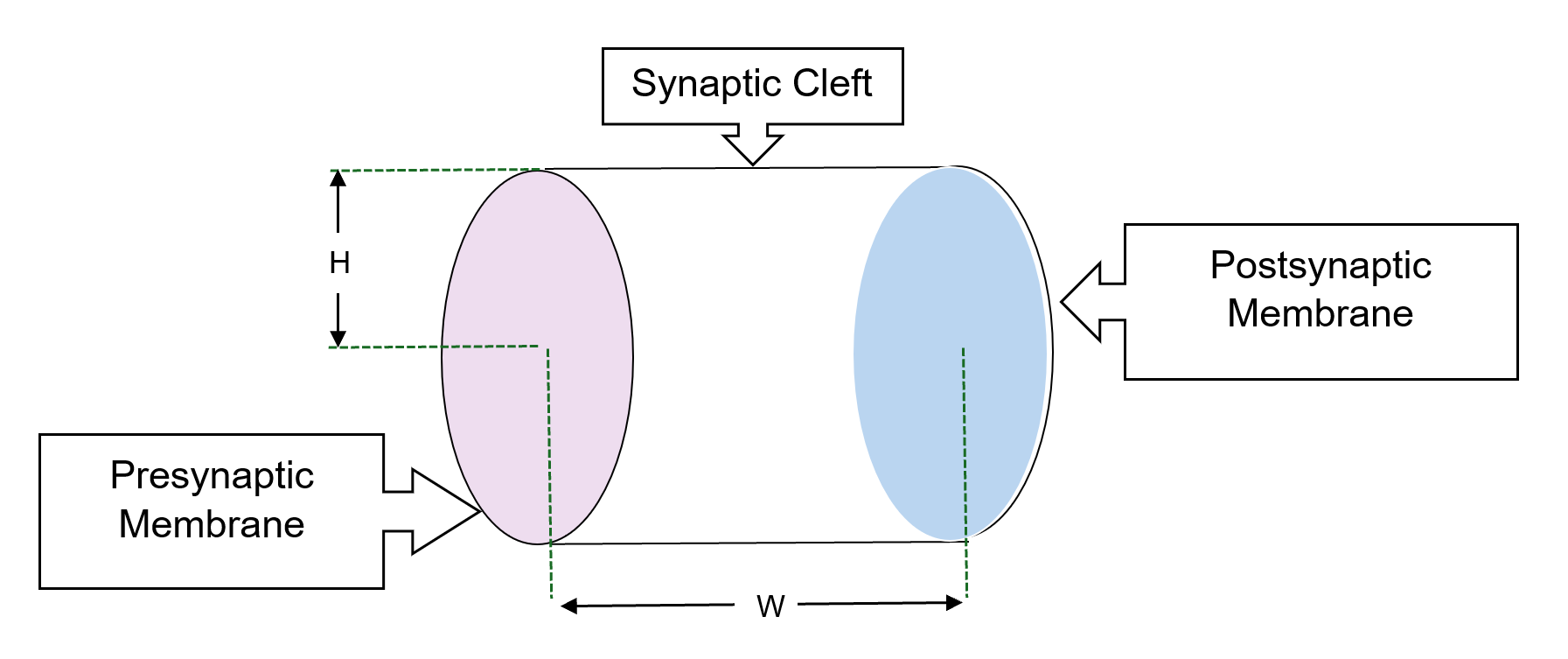}
\caption{Schematic of the synaptic cleft modeled as a cylinder.}
\label{variables}
\end{figure*}

In this study, we model the synaptic cleft as a cylindrical volume, as depicted in Fig. \ref{variables}, within which neurotransmitter diffusion occurs \cite{kulish2019modeling}.

\subsection{Molecular Communication Dynamics in the Synaptic Cleft}

Molecular communication (MC) represents an advanced approach focused on the transmission of information via the exchange of molecules \cite{akan2016fundamentals}. A key paradigm within MC is neuro-spike communication, which pertains to the signaling processes among neurons \cite{balevi2013physical, malak2013communication, ramezani2017information}. Within the synaptic cleft, the release of vesicles leads to interactions between glutamate and A$\beta$os, with A$\beta$os serving as diffusion obstacles. 

Research has shown that synaptic signaling is significantly disrupted in AD due to the presence of A$\beta$os; however, the precise molecular mechanisms underlying this disruption remain only partially understood \cite{tu2014oligomeric}. Exploring MC within this synaptic environment could provide valuable insights into potential therapeutic strategies.

Glutamate plays a critical role as a neurotransmitter in neuronal signaling. Its diffusion within the synaptic cleft, as well as the presence of obstacles along its path, directly influences the outcome of signaling—whether beneficial or pathological. Understanding how glutamate moves through the synaptic cleft and identifying any barriers it encounters during diffusion is essential for unraveling the mechanisms underlying various neuronal disorders.

\begin{figure}[ht]
\centering
\includegraphics[width=1\linewidth]{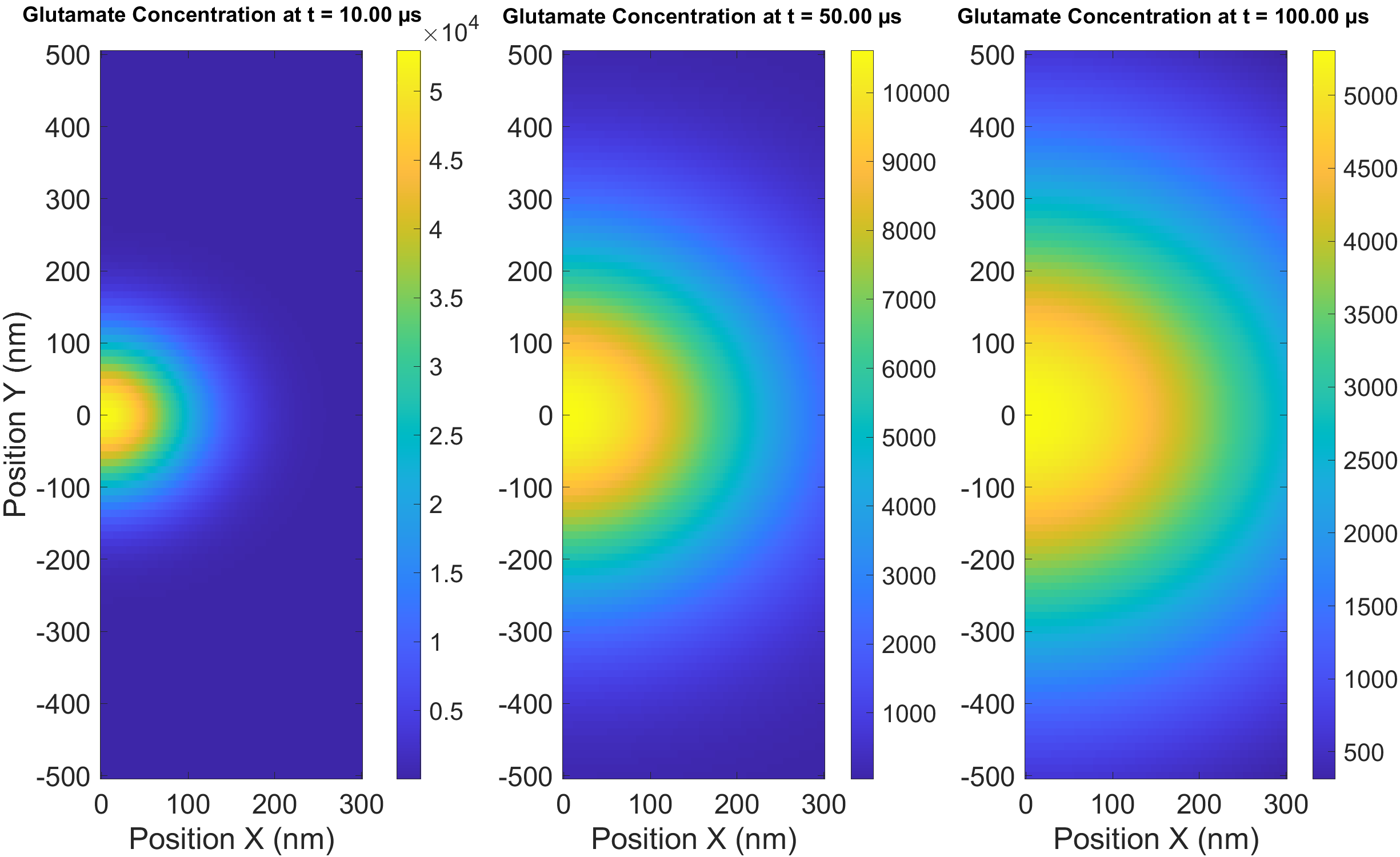}
\caption{Normal diffusion from a single vesicle containing approximately 2000 glutamate molecules  \cite{clements1992time}.}
\label{fig:first section}
\end{figure}

Glutamate is stored within vesicles and released into the synaptic cleft upon neuronal excitation \cite{hayashi2018structure}. Based on Fick's law of diffusion and the cylindrical model of the synaptic cleft presented in \cite{kulish2019modeling}, the normal diffusion of glutamate within the synaptic cleft is described as: 

\begin{equation}
\frac{\partial U(t, h, w)}{\partial t} = D \Delta U(t, h, w),
\label{equ: NormalDiff}
\end{equation}
where \( D \) represents the diffusion coefficient, and \( \Delta \) denotes the Laplace operator, indicating how the concentration \( U \) changes over time and space due to diffusion. The variable \( h \) denotes the distance from the cylinder axis to a point within the cleft, with a range of \( 0 \leq h \leq H \). Similarly, \( w \) represents the distance along the cylinder axis from the presynaptic membrane to a point within the system, with a range of \( 0 \leq w \leq W \). The normal diffusion and concentration of glutamate are illustrated in Fig. \ref{fig:first section}. However, obstacles significantly impact the diffusion process, with their characteristics, positioning, and movement playing crucial roles. The subsequent sections will explore these aspects in greater detail.

\section{System Model and Problem Formulation}
\label{section:III}
In this section, we first describe the characteristics of A$\beta$os, followed by an analysis of their stochastic movement, using experimental data summarized in Table \ref{table1}.
\subsection{Beta-Amyloid Oligomers Characteristics}
\label{section:III:A}
A$\beta$os are widely recognized as a critical factor in the progression of AD, contributing to neuronal damage and cognitive decline \cite{tolar2021neurotoxic, huang2020toxicity, kayed2013molecular, rinauro2024misfolded, du2022mitochondrial}. In the brains of individuals affected by AD, both fibrillar and prefibrillar forms of amyloid $\beta_{1-42}$ oligomers have been observed \cite{kayed2007fibril}. Prefibrillar oligomers exhibit greater mobility compared to their fibrillar counterparts. Despite being formed from different peptides, A$\beta$os share similar structural characteristics and display uniform diffusive behavior on the surface of living cells \cite{calamai2011single}.

A$\beta$os interact with various ions and proteins, leading to random forces that induce their stochastic movement and aggregation. Consequently, the presence of A$\beta$os in the synaptic cleft is inherently stochastic. This observation prompts the use of stochastic methods to model their presence effectively. 

\subsection{Stochastic Modeling of amyloid-beta}
\label{section:III:B}
Stochastic differential equations (SDEs) incorporate both stochastic (random) and deterministic components, as represented in (\ref{5}), where the functions \( b \) and \( a \) denote the stochastic and deterministic components, respectively. This distinguishes SDEs from ordinary differential equations (ODEs), which are purely deterministic. SDEs are particularly well-suited for systems influenced by random factors. The standard SDE formulation for modeling such systems is given as follows \cite{studysmarter_sde}:

\begin{equation}
dY = a(Y_t, t)dt + b(Y_t, t)dW_t.
\label{5}
\end{equation}

In addition, the SDE framework is extendable. Given the cylindrical symmetry of the synaptic cleft, a two-dimensional approach is necessary to model the stochastic presence of A$\beta$os. Therefore, we propose the following formulation:
\begin{equation}
\begin{aligned}
dX = & \, \left[ a(X_t, Y_t, t)dt + b_{11}(X_t, Y_t, t)dW_t \right. \\
     & \, \left. + b_{12}(X_t, Y_t, t)dZ_t \right],
\label{6}
\end{aligned}
\end{equation}
\begin{equation}
\begin{aligned}
dY = & \, \left[ a(X_t, Y_t, t)dt + b_{21}(X_t, Y_t, t)dW_t \right. \\
     & \, \left. + b_{22}(X_t, Y_t, t)dZ_t \right],
\label{7}
\end{aligned}
\end{equation}
where \( X_t \) and \( Y_t \) represent the components of a two-dimensional stochastic process, with \( dW_t \) and \( dZ_t \) as independent Wiener processes. The function \( a(X_t, Y_t, t) \) denotes the drift function that influences both the \( X \) and \( Y \) axes. The coefficients \( b_{11} \) and \( b_{22} \) are the diffusion coefficients along the \( X \) and \( Y \) axes, respectively. The coefficient \( b_{12} \) represents the diffusion along the \( X \) axis influenced by the Wiener process \( dZ_t \), while \( b_{21} \) accounts for the diffusion along the \( Y \) axis influenced by the Wiener process \( dW_t \). Together, these functions and coefficients define the dynamics of the stochastic process.

The first step in modeling the drift and diffusion functions is to monitor the movement of A$\beta$os. In this context, the mean square displacement (MSD) function is an effective tool for tracking the motion of atoms within large and complex structures \cite{riahi2019identifying}. MSD can be utilized to model the diffusion matrix in a SDE for A$\beta$os diffusion, as it is directly related to the diffusion coefficient 
(D), a critical component of the diffusion matrix.

Analysis of MSD and the initial D have been reported in the literature \cite{bannai2006imaging, ehrensperger2007multiple}. According to \cite{calamai2011single}, MSD is determined as:

 \begin{equation}
     MSD(n.dt) = \frac{1}{N-n} \sum_{i=1}^{N-n} \left[ (X_{i+n} - X_i)^2 + (Y_{i+n} - Y_i)^2 \right],
     \label{eq: MSD}
 \end{equation}
where $X_{i}$ and $Y_{i}$ denote a particle's coordinates in the $i_{\text{th}}$ frame, \textit{dt} is time intervals, \textit{N} is the trajectory's total frame number, and \textit{n} is time lag. This technique is crucial for analyzing particles' side-to-side movements, offering insights into both immediate (early D) and extended (movement types) behavior patterns. As the result of MSD calculation for both prefibrillar and fibrillar oligomers using (\ref{eq: MSD}) in \cite{calamai2011single}, we have prefibrillar oligomers D as  $4.0 \times 10^{-3} \mu m^2s^{-1}$ which is nearly four times more that fibrillar oligomers D which  is  $9.4 \times 10^{-4} \mu m ^2s^{-1}$. This finding from the literature allows for the formulation of the diffusion function matrix for prefibrillar oligomers as follows:
\[
B = \begin{bmatrix}
4 \times 10^{-3} & 0 \\
0 & 4 \times 10^{-3}
\end{bmatrix} \, \text{$\mu$m}^2/\text{s},
\]
it is inferred that D is isotropic, exhibiting uniform behavior along both the X and Y axes. This suggests that the coefficient remains constant over time, showing no variation with temporal changes.

The drift function represents the deterministic part of the particle movement. It captures how the system's behavior is influenced by its current state, including interactions with other molecules, local environmental conditions, or any factors specific to the system being modeled.

We model the drift function of amyloid stochastic presence in the synaptic cleft based on its root-mean-square velocity ($v_{\text{rms}}$), which is calculated as:

\[
v_{\text{rms}} = \sqrt{\frac{2dD}{\Delta t}} \, \text{$\mu$m/s},
\]
in the above $v_{\text{rms}}$ formula, $d$ represents the dimension of motion, $D$ denotes the diffusion coefficient, and $\Delta t$ refers to the time step.

Consequently, the SDE describing the movement of amyloid oligomers in the synaptic cleft can be defined as:

\begin{equation}
 \Delta X_t = -\sqrt{\frac{2\times2 \times 4\times 10^{-3} }{\Delta t}} \Delta t + 4 \times 10^{-3} \Delta W_t,
 \label{SDEX}
 \end{equation}
\begin{equation}
     \Delta Y_t = -\sqrt{\frac{2\times2 \times4\times 10^{-3} }{\Delta t}} \Delta t + 4 \times 10^{-3} \Delta Z_t.
      \label{SDEY}
\end{equation} 

It can be concluded that the stochastic movement of A$\beta$os indeed leads to stochastic distributions of these oligomers within the synaptic cleft.


\begin{table*}[t!]
    \centering
    \caption{Comparison of Glutamate and Amyloid Oligomer Characteristics}
    \begin{tabular*}{\textwidth}{@{\extracolsep{\fill}} l l l}
        \hline
        \textbf{Characteristic} & \textbf{Glutamate} & \textbf{Amyloid Oligomer} \\
        \hline
      Radii & $\approx 0.5\,\text{\AA}$ & $7.5\,\text{nm} < R < 17.5\,\text{nm}$ \cite{chimon2007evidence} \\
        \hline
        Mass & 147.13 daltons (g/mol) & $> 50\,\text{kDa}$ \cite{lacor2007abeta}\cite{velasco2012synapse} \\
        \hline
        Diffusion Coefficient & 300 $\mu\text{m}^2 \text{s}^{-1}$ \cite{li2022computational}\cite{wathey1979numerical} & $4.0 \times 10^{-3} \mu\text{m}^2 \text{s}^{-1}$ \cite{calamai2011single} \\
        \hline
        $v_{\text{rms}}$ & $\approx \frac{37}{\Delta t} \, \mu\text{m} \text{s}^{-1}$ & $\approx \frac{0.12}{\Delta t} \, \mu\text{m} \text{s}^{-1}$ \\
        \hline
    \end{tabular*}
    \label{table1}
\end{table*}

\section{Interaction Between Glutamate and Beta-Amyloid Oligomers}
\label{section:IV}
In this section, first we analyze the dynamics of collisions; next, we explore the frequency of these collisions; and finally, we examine the consequences of these interactions within the channel.

\begin{figure}[ht]
\centering
\includegraphics[width=1\linewidth]{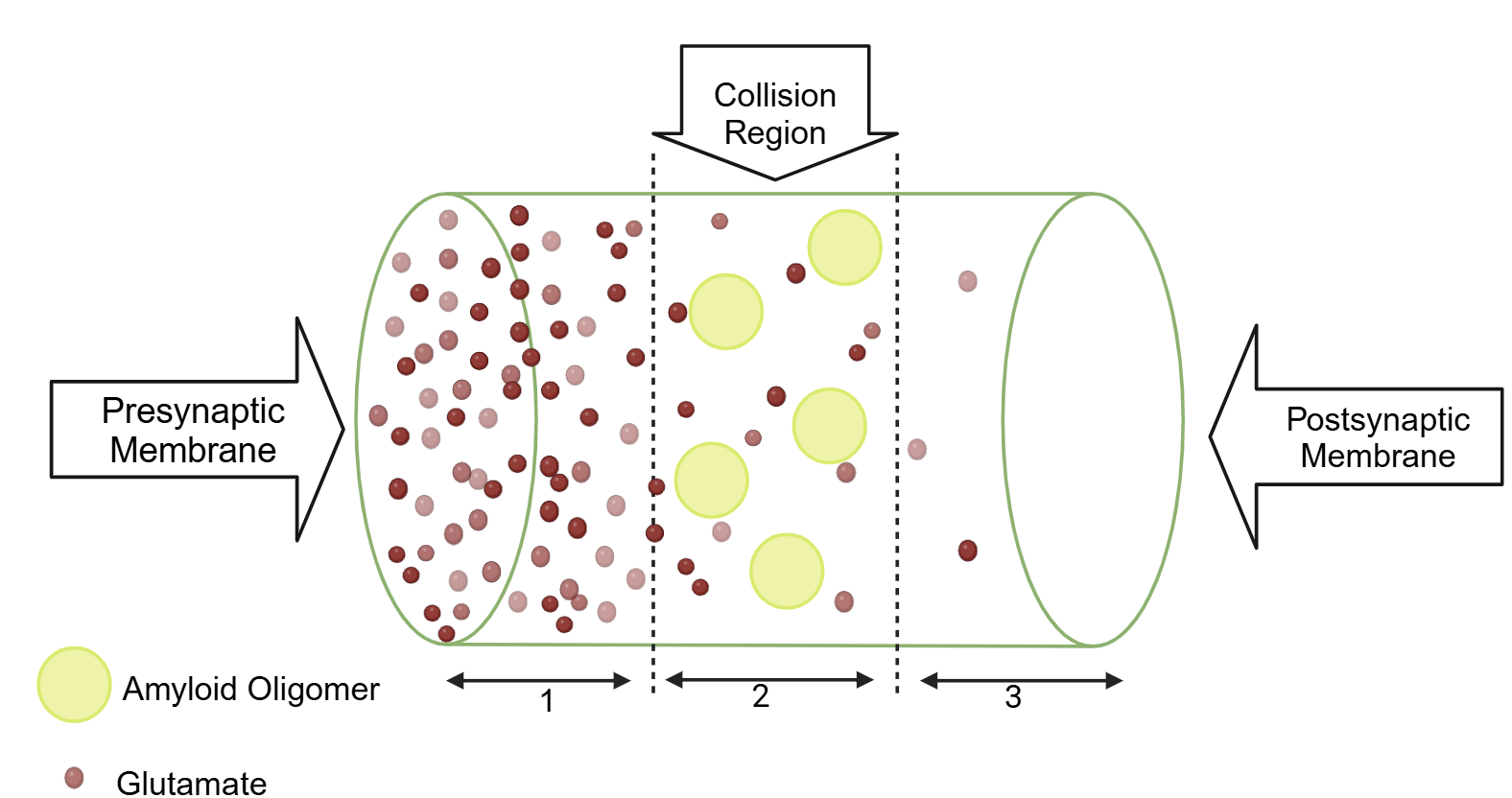}
\caption{
In the initial region, glutamate diffuses normally. In region 2, collisions with A$\beta$os disrupt its usual distribution.
}
\label{Regions}
\end{figure}

According to the findings presented in \cite{vilaseca2011new}, a random distribution of obstacles significantly impedes the diffusion of particles more than a normal distribution does, as it hinders the particles' ability to predict their paths effectively. Similarly, the random presence of A$\beta$os within the synaptic cleft causes random collisions with glutamate. From the moment glutamate starts colliding with A$\beta$os, it initiates Brownian motion, a process where particles move unpredictably due to collisions, as discussed in \cite{meyer2011particle}. In this study, as illustrated in Fig. \ref{Regions}, we divide the synaptic cleft, the channel for neuronal transmission, into three main regions: the first region, where glutamate undergoes normal diffusion; the second region, where collisions occur; and the third region, where particle movements are influenced by the collisions in the second region.

\subsection{Collision Dynamic}
\label{section:IV:A}
\begin{figure}[ht]
\centering
\includegraphics[width=0.5\textwidth]{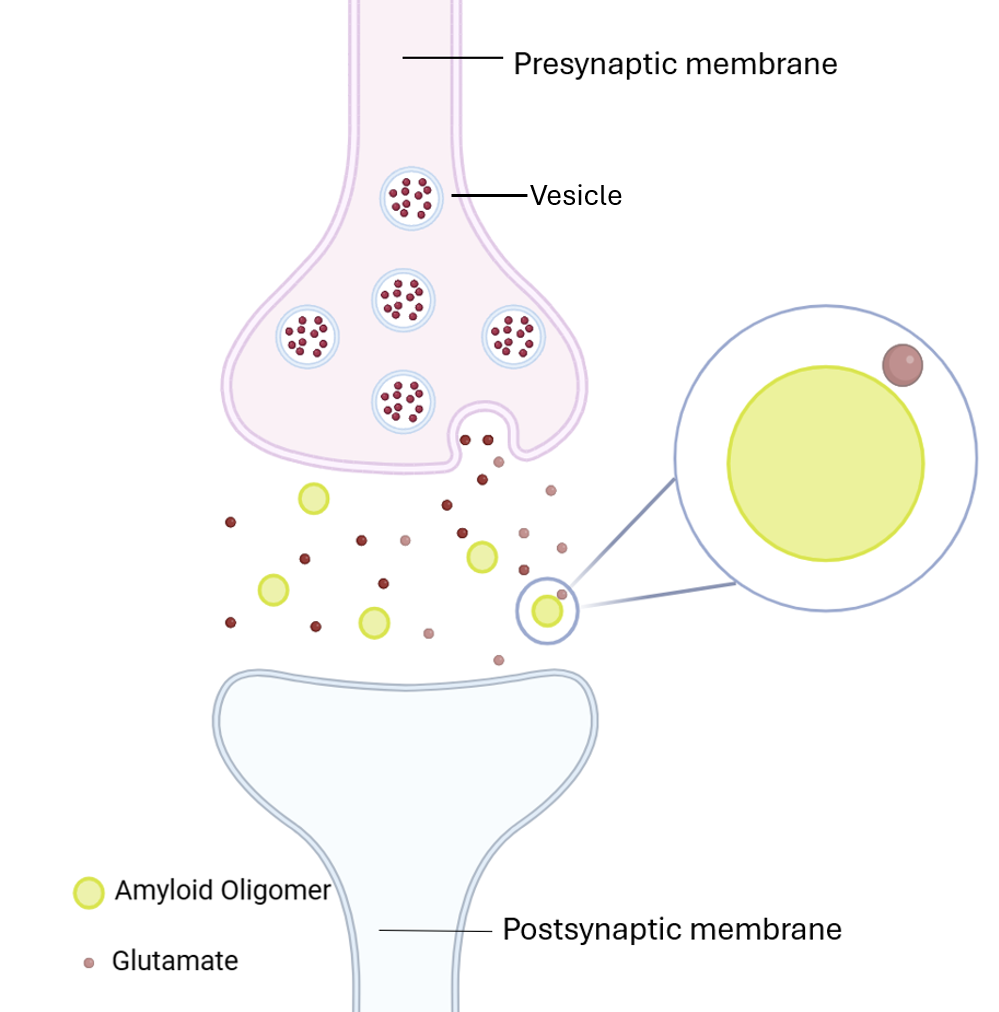}
\caption{A\bm$\beta$os as obstacles for glutamate diffusion.}
\label{Fig: obstacleandglutamate}
\end{figure}
Fig. \ref{Fig: obstacleandglutamate} illustrates the collisions between glutamate and A\bm$\beta$os. It is evident that the collision between A$\beta$os and glutamate adheres to the principle of momentum conservation, as their respective momenta prior to the collision are approximately equal. Under this assumption, the conservation of momentum for the two entities can be expressed as:
\begin{equation}
    m_A \cdot v_{Ax1} + m_B \cdot v_{Bx1} = m_A \cdot v_{Ax2} + m_B \cdot v_{Bx2},
    \label{eq: momentum}
\end{equation}
where \( m_A \) and \( m_B \) represent the masses of two particles. \( v_{Ax1} \) and \( v_{Bx1} \) denote the velocities of particles \( A \) and \( B \) before the collision, respectively, while \( v_{Ax2} \) and \( v_{Bx2} \) are their velocities after the collision.
To analyze the nature of collisions and the behavior of particles after impact, it is essential to determine the coefficient of restitution, denoted as \(e\). This coefficient represents the ratio of the relative velocity of the particles after the collision to their relative velocity before the collision, measured along the line of impact \cite{burko2019two}. The value of \(e\) ranges between 0 and 1. When \(e = 1\), the collision is elastic, meaning no kinetic energy is lost. Conversely, when \(e = 0\), the collision is perfectly inelastic, causing the objects to stick together and move as one after the collision, resulting in the maximum possible loss of kinetic energy within the system. Values of \(e\) between 0 and 1 represent partially elastic collisions, where some kinetic energy is lost. \(e\) is defined as:

\begin{equation}
    e = \frac{|v_{B,x,2} - v_{A,x,2}|}{|v_{B,x,1} - v_{A,x,1}|}.
    \label{eq: elasticity}
\end{equation}

\begin{figure}[ht]
\centering
\includegraphics[width=0.9\linewidth]{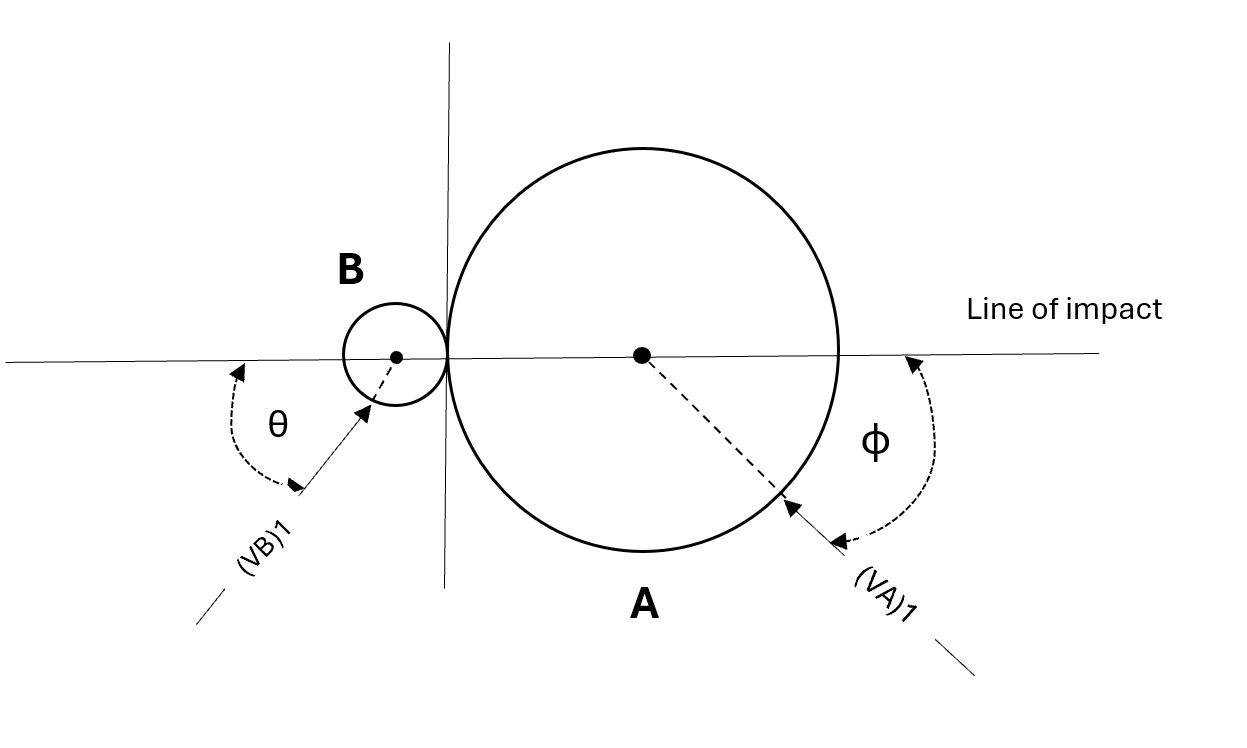}
\caption{
Particles A (A$\beta$o) and B (glutamate) collide, possessing angular relative velocities denoted by angles \(\theta\) and \(\phi\) with respect to the line of impact.
}
\label{impact}
\end{figure}

\subsection{Collision Frequency} 
\label{section:IV:B}
The frequency of collisions between two particles undergoing Brownian motion is calculated as follows \cite{meyer2011particle}:
\begin{equation}
    F = \left( \frac{2k_B T}{3\mu} \right) \left( \frac{1}{r} + \frac{1}{R} \right) (r + R),
    \label{frequency}
\end{equation}
where \(k_B\) is the Boltzmann constant, \(T\) represents the absolute temperature, \(\mu\) denotes the fluid viscosity, and \(r\) and \(R\) are the radii of the particles. (\ref{frequency}) is based on the following assumptions:
\begin{enumerate}
    \item The collision efficiency is defined as one. 
    \item The fluid motion is laminar.
    \item All the particles are the same size, a condition known as monodispersity.
    \item Breaking of particles is not considered in this scenario.
    \item The particles are spherical and will remain the same after collision.
    \item Only two particles are involved in the collision.
\end{enumerate}
In the analysis of A$\beta$o and glutamate collisions, we apply the previously mentioned assumptions for collision frequency, with the exception of collision efficiency ($\alpha$), which can range between 0 and 1. As a result, the number of successful collisions differs depending on the value of $\alpha$.  Consequently, we propose that the collision frequency can be given by:

\begin{equation}
    F = \alpha \left( \frac{2k_B T}{3\mu} \right) \left( \frac{1}{r} + \frac{1}{R} \right) (r + R).
     \label{equ:Frequency2}
\end{equation}

\subsection{Effects of Amyloid Beta Oligomers and Glutamate Collisions}
\label{section:IV:C}

\subsubsection{Impact on Glutamate Diffusion}
\label{section:IV:C:1}
The stochastic presence of A$\beta$os within the synaptic cleft significantly influences glutamate diffusion and uptake. The obstacles posed by both the size and concentration of A$\beta$os impede the mobility of glutamate molecules. The effective diffusion coefficient, $\tilde{D}_{\text{eff}}$, can be modeled by the equation presented in \cite{novak2011diffusion}:

\begin{equation}
    \tilde{D}_{\text{eff}}(\phi) \approx \left(1 - \frac{\phi}{\phi_c}\right)^{\lambda \phi_c},
    \label{equ:Diff}
\end{equation}
where $\phi$ represents the excluded volume fraction, indicating the volume occupied by the obstacles. The parameter $\phi_c$ denotes the percolation threshold, which is the critical volume fraction at which diffusion is severely hindered. The parameter $\lambda$ relates to the geometrical configuration and shape of the obstacles.

\begin{figure}[ht]
\centering
\includegraphics[width=1.00\linewidth]{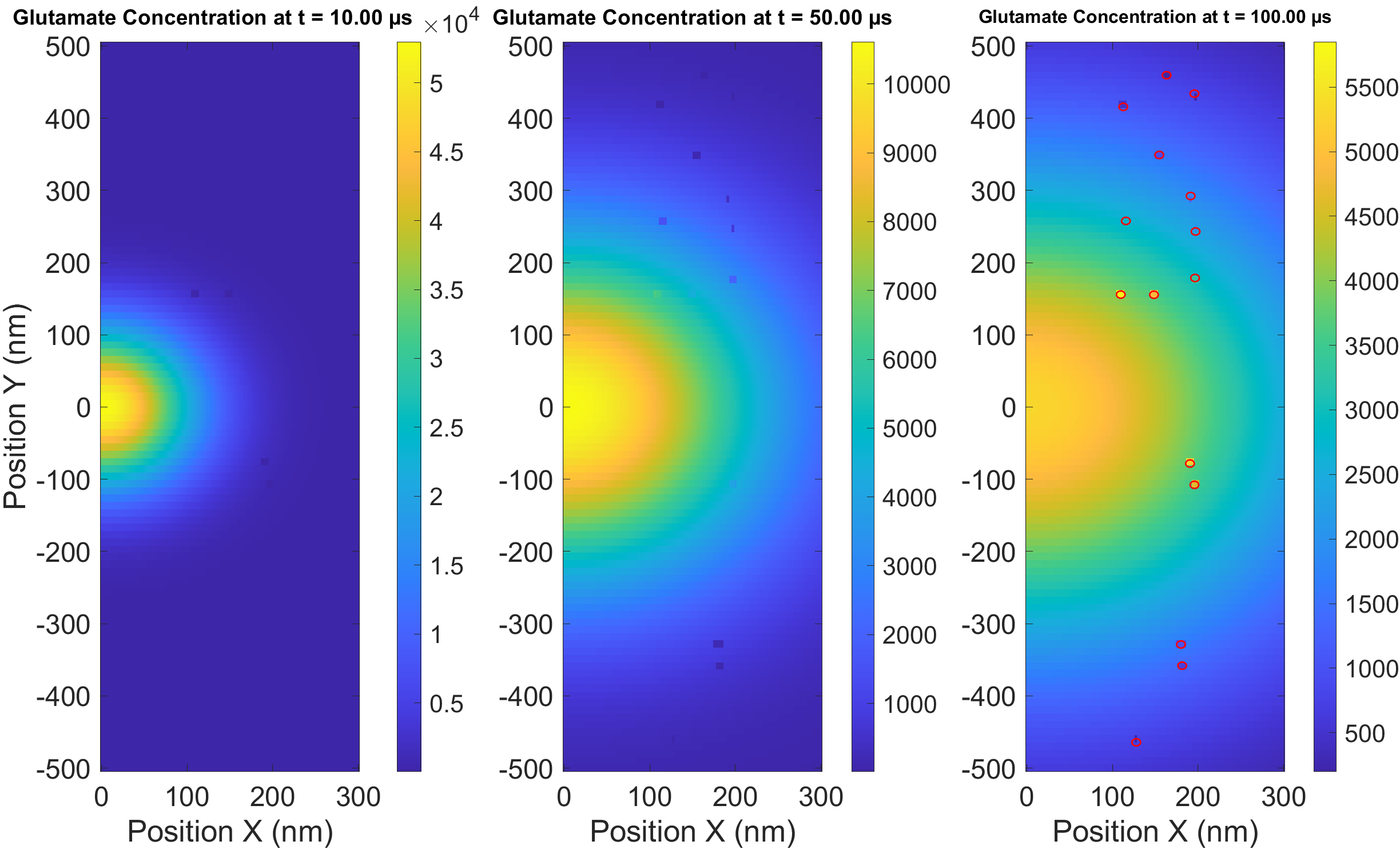}
\caption{Impact of A$\beta$o presence and distribution on glutamate diffusion over time.}
\label{Fig:finalconcentration}
\end{figure}

Fig. \ref{Fig:finalconcentration} visually demonstrates the impact of A$\beta$o presence and distribution on glutamate diffusion over time. Initially, glutamate spreads relatively uniformly; however, as time progresses and A$\beta$os are introduced, the diffusion pattern becomes disrupted, resulting in localized regions of reduced glutamate concentration corresponding to the locations of A$\beta$os. This visualization reinforces the conclusion that A$\beta$os can significantly hinder glutamate diffusion within the synaptic cleft.

\subsubsection{Impact on Beta-Amyloid Oligomer Movement}
\label{section:IV:C:2}
When a collision occurs between glutamate and A$\beta$o, the resulting force impels the A$\beta$o in the direction of glutamate's initial trajectory, typically towards the postsynaptic membrane. This directional movement is driven by the transfer of momentum during the collision.

The kinematic equation \(x = vt\) can be employed to estimate the displacement of A$\beta$os after the collision, where \(v\) is the velocity imparted to the oligomer, and \(t\) is the time elapsed. This equation provides a simplified model to understand how the A$\beta$o moves toward the postsynaptic site following its interaction with glutamate.

\subsubsection{Impact on NMDAR Dysregulation}
\label{section:IV:C:3}
NMDARs are essential for synaptic transmission, plasticity, and the functional maintenance of the nervous system. However, their dysfunction is associated with neurotoxicity, seizures, ischemic stroke, and neurodegenerative diseases such as AD \cite{liu2019role, paoletti2013nmda, kodis2018n, macdermott1986nmda, muller2009both, villmann2007hypes, benarroch2011nmda, wenk2006neuropathologic}. Activation of NMDARs leads to an increase in cytosolic free intracellular calcium (\([Ca^{2+}]\)), a crucial factor for Long Term Potentiation (LTP) \cite{macdermott1986nmda, muller2009both}. 

\begin{figure}[ht]
\centering
\includegraphics[width=0.6\linewidth]{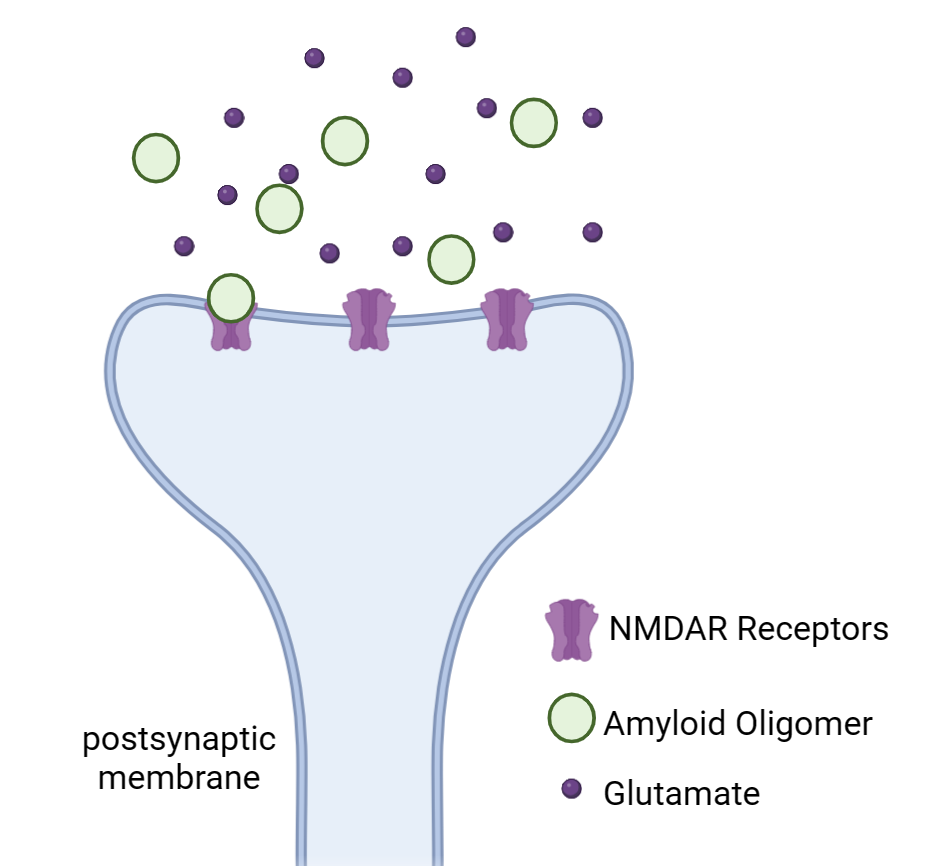}
\caption{Binding of A$\beta$o to NMDARs prevents them from receiving released neurotransmitters, adversely affecting LTP.}
\label{fig:NMDARs}
\end{figure}

Furthermore, emerging evidence increasingly suggests that A$\beta$os disrupt glutamate receptor function, leading to disturbances in glutamatergic synaptic transmission, which are closely linked to early cognitive deficits. NMDARs play a central role in the pathophysiology associated with A$\beta$os for several key reasons. First, NMDAR function is directly impacted by A$\beta$os, making it a critical target. Second, NMDARs are essential mediators of the effects of A$\beta$ on synaptic transmission and plasticity. Third, NMDARs may serve as receptors for A$\beta$os, either through direct or indirect interactions. Finally, NMDAR activity could modulate A$\beta$ production itself \cite{malinow2012new}. In vivo studies suggest a regulatory relationship where lower NMDAR activation is correlated with increased A$\beta$ production, while higher activation levels are associated with a reduction \cite{verges2011opposing}. Consequently, as depicted in Fig. \ref{fig:NMDARs}, the binding of A$\beta$os to NMDARs impairs the receptors' ability to receive released neurotransmitters, thus negatively impacting LTP.

$\bullet$ Kinetic Model for A$\beta$os Binding to NMDARs

 we can propose the A$\beta$os binding rate to NMDARs by applying the principles of mass-action kinetics, the rate of a first-order reaction \(A \rightarrow C\) is given by \(k[A]\), and the rate of a second-order reaction \(A + B \rightarrow C\) is \(k[A][B]\), where \([X]\) represents the concentration (or the number of molecules, adjusted for volume) of species \(X\) \cite{resat2009kinetic}.

The binding rate constant \( k \) is assumed to vary inversely with the size of the amyloid oligomer:
\[
k = k_{\text{base}} \times \left(\frac{50}{\text{size}}\right),
\]
where \( k_{\text{base}} \) is a base rate constant chosen arbitrarily. The factor \( \frac{50}{\text{size}} \) accounts for the hypothesis that larger oligomers have slower diffusion rates and greater steric hindrance, making it more difficult for them to move within the cleft and bind to receptors.

The differential equations based on these reaction rates are:
\begin{itemize}
    \item $\frac{d[A]}{dt} = -k[A][R]$ for A$\beta$os,
    \item $\frac{d[R]}{dt} = -k[A][R]$ for NMDA receptors,
    \item $\frac{d[AR]}{dt} = k[A][R]$ for the complex.
\end{itemize}

These equations collectively describe a bimolecular reaction where A$\beta$os and NMDA receptors bind to form a complex, following second-order kinetics with a rate constant k. The formation of the complex decreases the concentrations of both free A$\beta$os and NMDA receptors, and the stoichiometry of the reaction is 1:1, assuming no other pathways or reactions are involved.

\subsubsection{Impact on Signal-to-Noise Ratio (SNR)}
\label{section:IV:D}

MC allows implantable devices to transmit information using molecules as carriers. However, a major challenge in MC is molecular noise, which increases the likelihood of communication errors. This noise is particularly relevant in pathological conditions like AD, where cellular disruptions lead to an imbalance in molecular processes, making affected cells more reactive and thus noisier \cite{borges2024cell}. The SNR is a critical metric for evaluating the efficiency of MC within the synaptic cleft, especially in the presence of A$\beta$os. In this context, the signal refers to the successful diffusion and binding of glutamate molecules to postsynaptic receptors, while the noise is represented by the collisions with A$\beta$os, which impede this diffusion process. The SNR quantitatively measures the ability to distinguish the signal from the noise, thereby illustrating the impact of A$\beta$os on synaptic transmission.

To calculate the SNR, the signal power is modeled as the MSD of glutamate in an obstacle-free environment, representing optimal neurotransmitter diffusion. In contrast, the noise power is derived from the variance in glutamate diffusion due to the stochastic presence of A$\beta$os, which act as physical barriers within the synaptic cleft. The SNR is then expressed as the ratio of signal power to noise power, providing insights into the extent to which A$\beta$os impair glutamate transmission.

The SNR can thus be formulated as:
\[
\text{SNR} = \frac{\text{MSD}_g \text{ (without obstacles)}}{\text{MSD}_g \text{ (with obstacles)} - \text{MSD}_g \text{ (without obstacles)}},
\]
where $\text{MSD}_g$ represents the mean squared displacement of glutamate.

\section{Results and discussion}
\label{section:V}
In this section, we discuss the results of our numerical simulations analyzing the interactions between glutamate and A$\beta$os within the synaptic cleft. These simulations were conducted using a mathematical model developed in MATLAB, which builds upon the theoretical framework outlined in previous chapters. Our study focuses on evaluating how the size and concentration of A$\beta$os affect several key parameters: collision frequency, glutamate diffusion and its MSD, movement of A$\beta$os towards the postsynaptic membrane, their binding rate with NMDARs, and the SNR of the synaptic channel. Additionally, we examine the implications of these interactions for synaptic dysfunction, particularly in the context of AD. 

\begin{figure}[ht!]
  \centering
  \begin{subfigure}[b]{0.45\textwidth}
    \includegraphics[width=\textwidth]{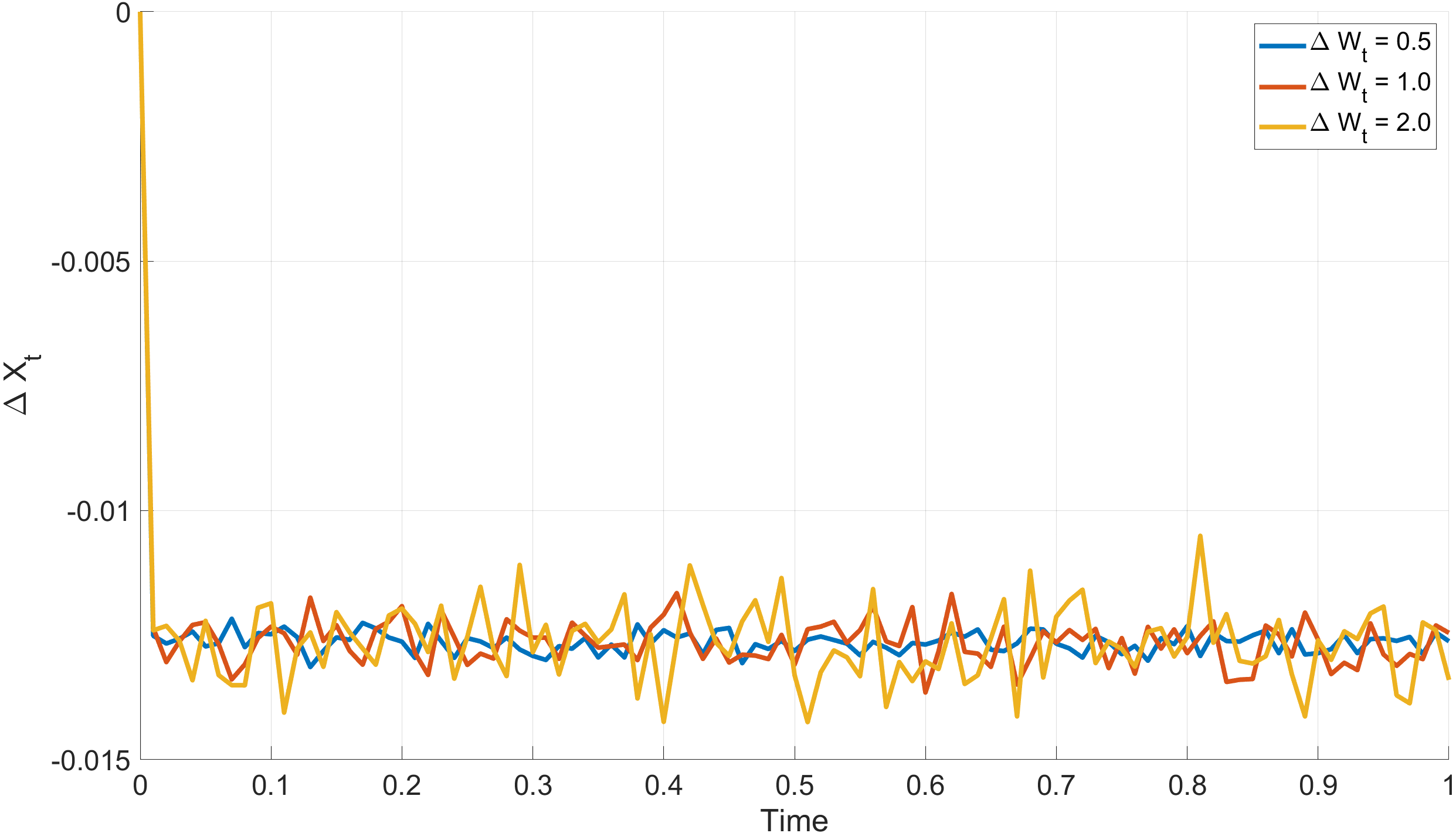}
    \caption{}
    \label{fig: X-axis}
  \end{subfigure}
  \hfill 
  \begin{subfigure}[b]{0.45\textwidth}
    \includegraphics[width=\textwidth]{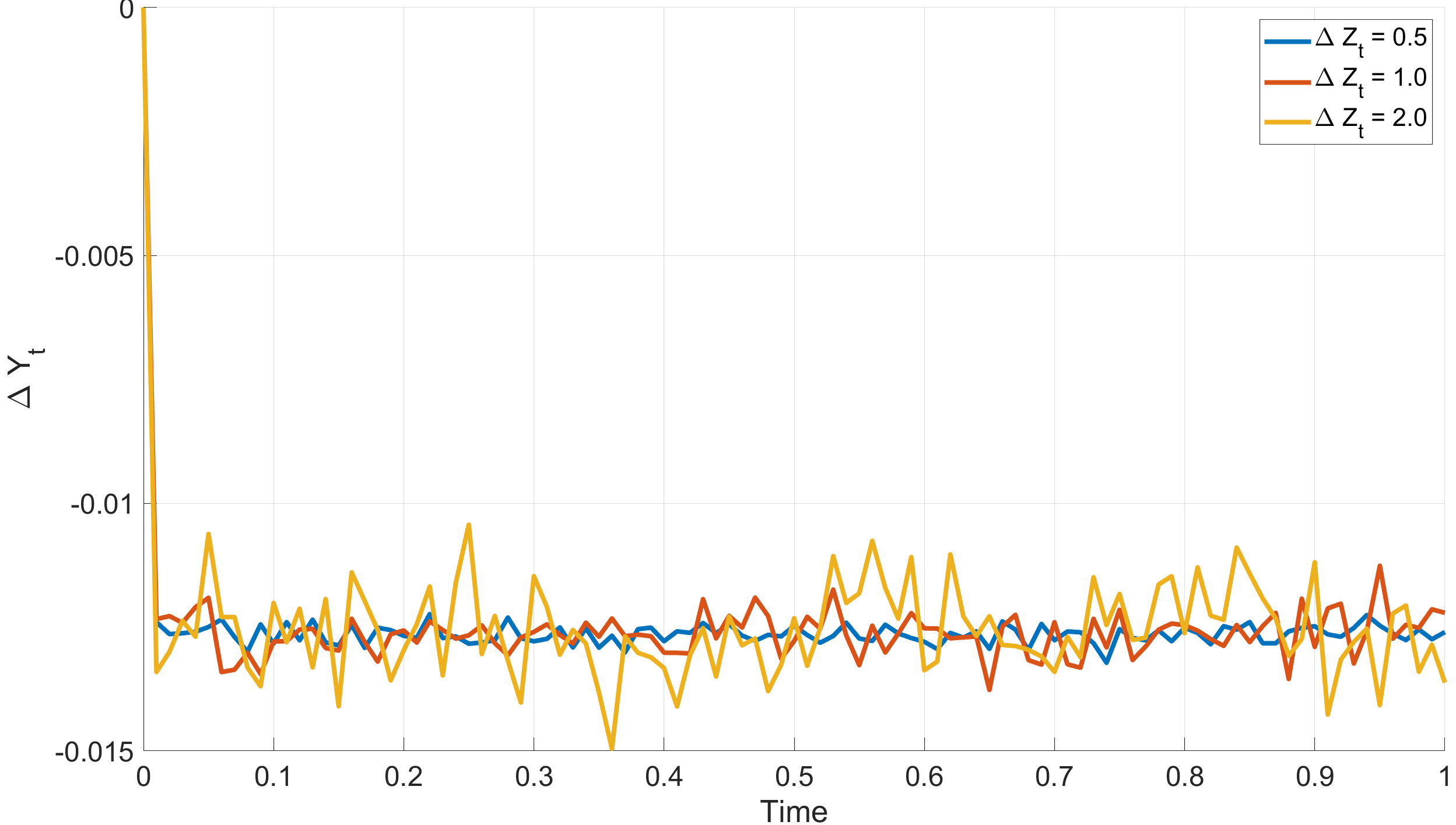}
    \caption{}
    \label{fig: Y-Axis}
  \end{subfigure}
  \caption{Stochastic distribution of A$\beta$os within the synaptic cleft, represented along the X-axis (a) and along the Y-axis (b) with variations influenced by different Wiener processes.}
  \label{fig: SDE}
\end{figure}

To initiate the analysis, we reference the stochastic differential equations (\ref{SDEX}) and (\ref{SDEY}) introduced in Sec. \ref{section:III:B}, which describe the random movement of A$\beta$os within the synaptic cleft. Fig. \ref{fig: SDE} illustrates the stochastic distribution of A$\beta$os, highlighting three distinct scales of the Wiener process. This figure demonstrates how variations in the Wiener process significantly impact the random motion, leading to diverse positioning of A$\beta$os within the cleft.

\subsection{Collision Frequency}
\begin{figure}[ht]
\centering
\includegraphics[width=1\linewidth]{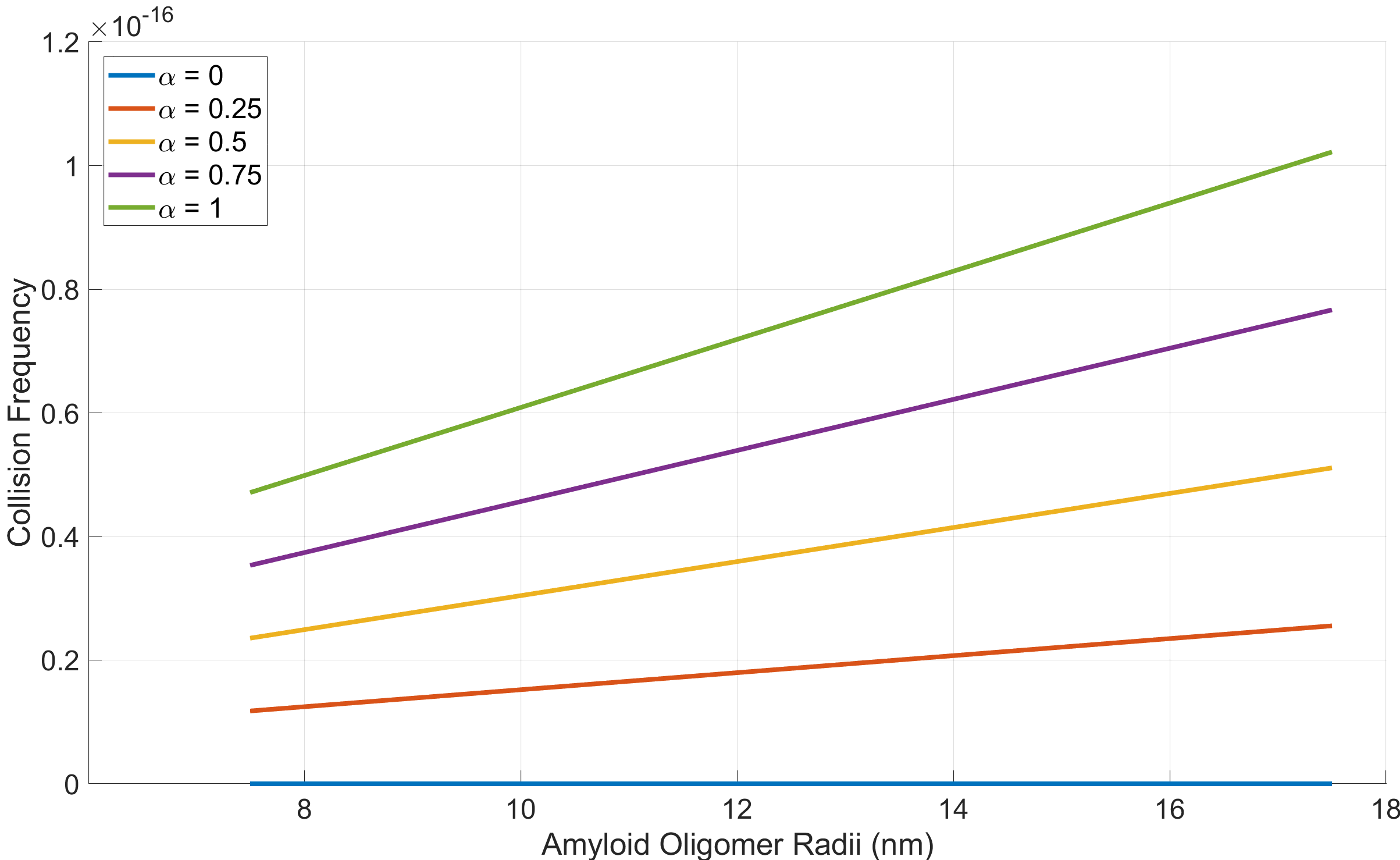}
\caption{Collision frequency of A$\beta$o and glutamate for varying A$\beta$o radii.}
\label{fig: frequency}
\end{figure}

In Sec. \ref{section:IV:B}, the collision frequency has been analyzed, with the corresponding simulation results presented in Fig. \ref{fig: frequency}. The parameters used for the simulation are detailed in Table \ref{tab:simulation_parameters}. The results depict the collision frequency between A$\beta$o and glutamate as a function of A$\beta$o size, considering five different collision efficiencies. The data indicate that as the size of A$\beta$o increases, the collision frequency also rises linearly. Moreover, the analysis suggests that higher collision efficiency ($\alpha$) results in a proportionately higher collision frequency, with $\alpha = 1$ yielding the highest frequency across all oligomer sizes.

Notably, for $\alpha = 0$, the collision frequency remains zero, regardless of the A$\beta$o size, representing a boundary condition where no collisions occur. These findings underscore the importance of considering both the size of A$\beta$o and the efficiency of collisions in evaluating their potential impact on synaptic function.

\begin{figure}
    \centering
\end{figure}

\subsection{Glutamate Diffusion}
\begin{figure}[ht]
    \centering
    \begin{subfigure}[b]{0.5\textwidth}
        \centering
        \includegraphics[width=\textwidth]{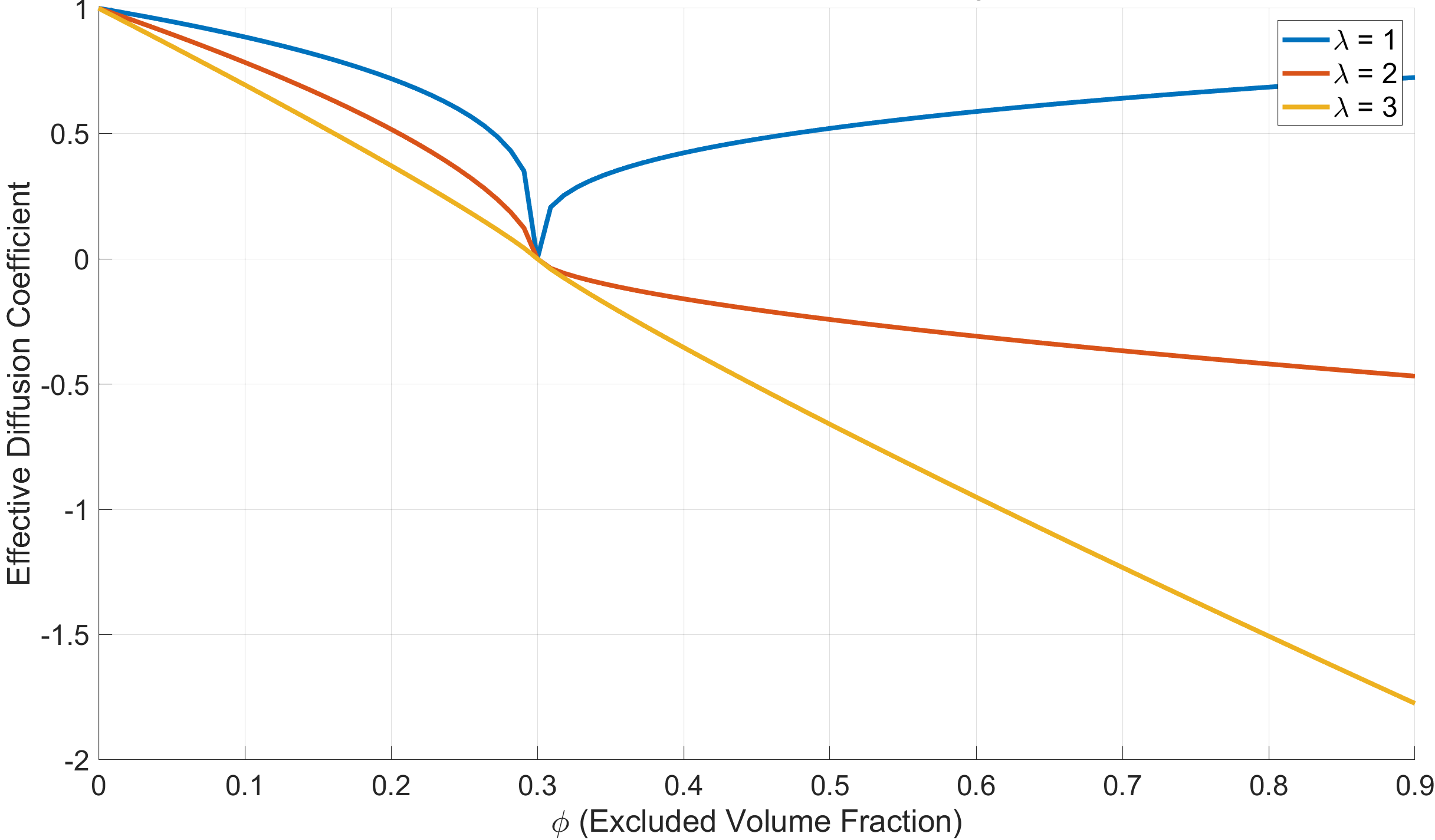} 
        \caption{}
        \label{Fig: Effective Diffusion_a}
    \end{subfigure}
    \hfill
    \begin{subfigure}[b]{0.5\textwidth}
        \centering
        \includegraphics[width=\textwidth]{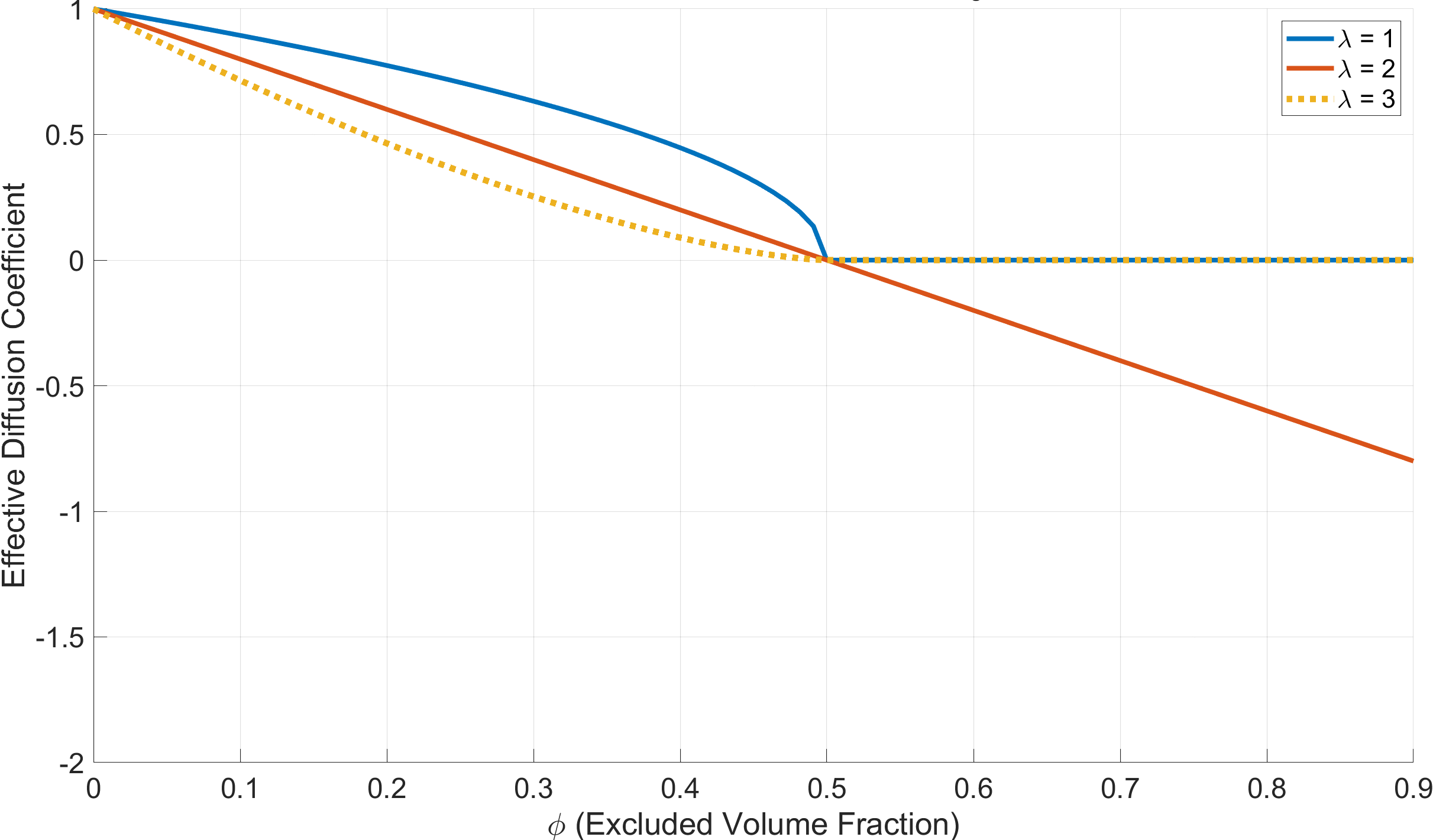} 
        \caption{}
        \label{Fig: Effective Diffusion_b}
    \end{subfigure}
    \hfill
    \begin{subfigure}[b]{0.5\textwidth}
        \centering
        \includegraphics[width=\textwidth]{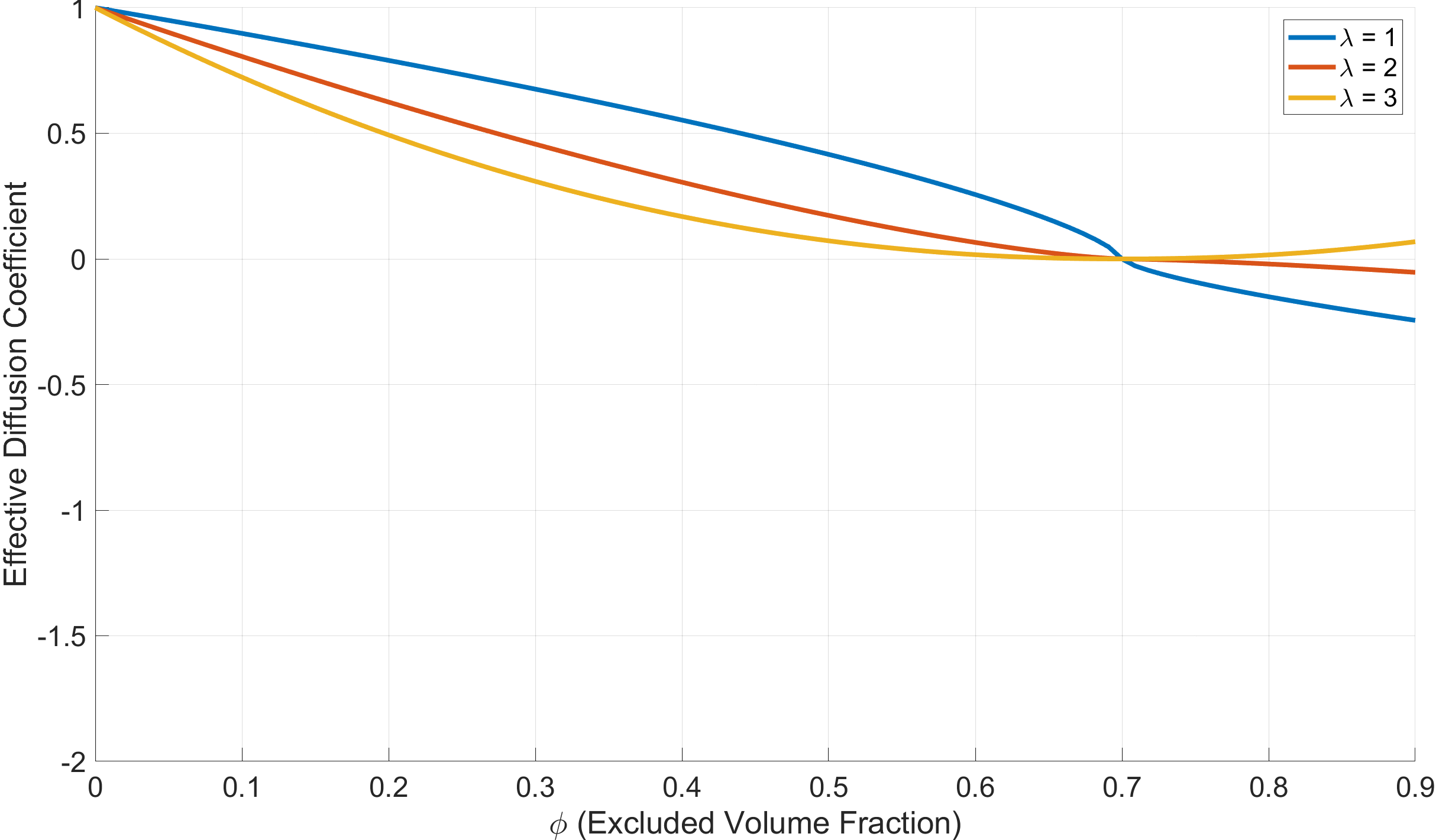} 
        \caption{}
        \label{Fig: Effective Diffusion_c}
    \end{subfigure}
    
   \caption{Effective Diffusion Coefficient of Glutamate for (a) $\phi_c = 0.3$, (b) $\phi_c = 0.5$, and (c) $\phi_c = 0.7$, with $\lambda = 1, 2, 3$.}

    \label{Fig: Effective Diffusion}
\end{figure}

In Sec. \ref{section:IV:C:1}, we considered (\ref{equ:Diff}) to study the effect of A$\beta$os concentration and size on glutamate diffusion. As shown in Fig. \ref{Fig: Effective Diffusion}, the graphs provided depict the effective diffusion coefficient of glutamate as a function of the excluded volume fraction ($\phi$) occupied by amyloid obstacles, considering different values of the percolation threshold ($\phi_c$) and a parameter $\lambda$, which relates to the geometrical configuration and shape of the obstacles. For Fig. \ref{Fig: Effective Diffusion_a} with $\phi_c = 0.3$, it is observed that for $\lambda = 1$, the effective diffusion coefficient decreases rapidly as $\phi$ increases, with a slight recovery after a certain point, suggesting a complex interaction between glutamate molecules and amyloid obstacles. For $\lambda = 2$ and $\lambda = 3$, the diffusion coefficient decreases steadily without recovery, indicating that a higher $\lambda$ leads to a more substantial reduction in diffusion. This suggests that when $\phi_c$ is low, indicating a lower percolation threshold, the diffusion of glutamate is significantly hindered even at smaller excluded volume fractions. The effect is more pronounced for higher $\lambda$, implying that as the complexity of the obstacle geometry increases, the hindrance to glutamate diffusion also increases. If we consider the size of A$\beta$o as an influential parameter on $\lambda$, we can conclude that larger A$\beta$o sizes have a more significant negative impact on glutamate diffusion compared to smaller ones.

For Fig. \ref{Fig: Effective Diffusion_b}, with $\phi_c = 0.5$, it is observed that for $\lambda = 1$, the diffusion coefficient decreases smoothly yet significantly. For $\lambda = 2$ and $\lambda = 3$, the curves exhibit a more complex behavior, with a noticeable change in slope around $\phi = 0.5$, particularly for $\lambda = 2$. This indicates that with a higher percolation threshold of $\phi_c = 0.5$, the initial diffusion of glutamate is less hindered compared to Fig. \ref{Fig: Effective Diffusion_a}. However, as $\phi$ increases, the hindrance becomes more pronounced, especially for $\lambda = 2$ and $\lambda = 3$. This suggests that the geometrical configuration and shape of the obstacles play a critical role in determining diffusion behavior when the obstacles occupy a larger volume.

In Fig. \ref{Fig: Effective Diffusion_c}, with $\phi_c = 0.7$, it is observed that for all values of $\lambda$, the effective diffusion coefficient decreases steadily as $\phi$ increases, with no recovery observed as in Fig. \ref{Fig: Effective Diffusion_a}. The decrease is more gradual compared to Figs. \ref{Fig: Effective Diffusion_a} and \ref{Fig: Effective Diffusion_b}, particularly for lower values of $\lambda$. With the highest percolation threshold of $\phi_c = 0.7$, the diffusion of glutamate is less affected by lower $\phi$ values, indicating that obstacles must occupy a larger volume fraction before significantly hindering diffusion. The steady decrease across all $\lambda$ values suggests that at higher $\phi_c$, the shape and geometrical configuration of the obstacles have a less pronounced effect on glutamate diffusion compared to lower $\phi_c$ values.

In conclusion, both the concentration and size (as reflected by $\phi$ and $\lambda$) of amyloid obstacles significantly influence glutamate diffusion in the synaptic cleft. Lower percolation thresholds ($\phi_c$) result in more substantial hindrance to diffusion, particularly at smaller excluded volume fractions, whereas higher thresholds require a more significant volume fraction to produce a similar effect. The parameter $\lambda$, which reflects the complexity of the obstacles' geometry, plays a crucial role in determining the rate at which the diffusion coefficient decreases with increasing $\phi$.  This result aligns with the findings of \cite{vilaseca2011new}, which indicate that as the number of obstacles increases, the dispersion of particles becomes more gradual.

\subsection{Glutamate MSD}
\begin{figure}[ht]
  \centering
  \begin{subfigure}[b]{0.45\textwidth}
    \includegraphics[width=\textwidth]{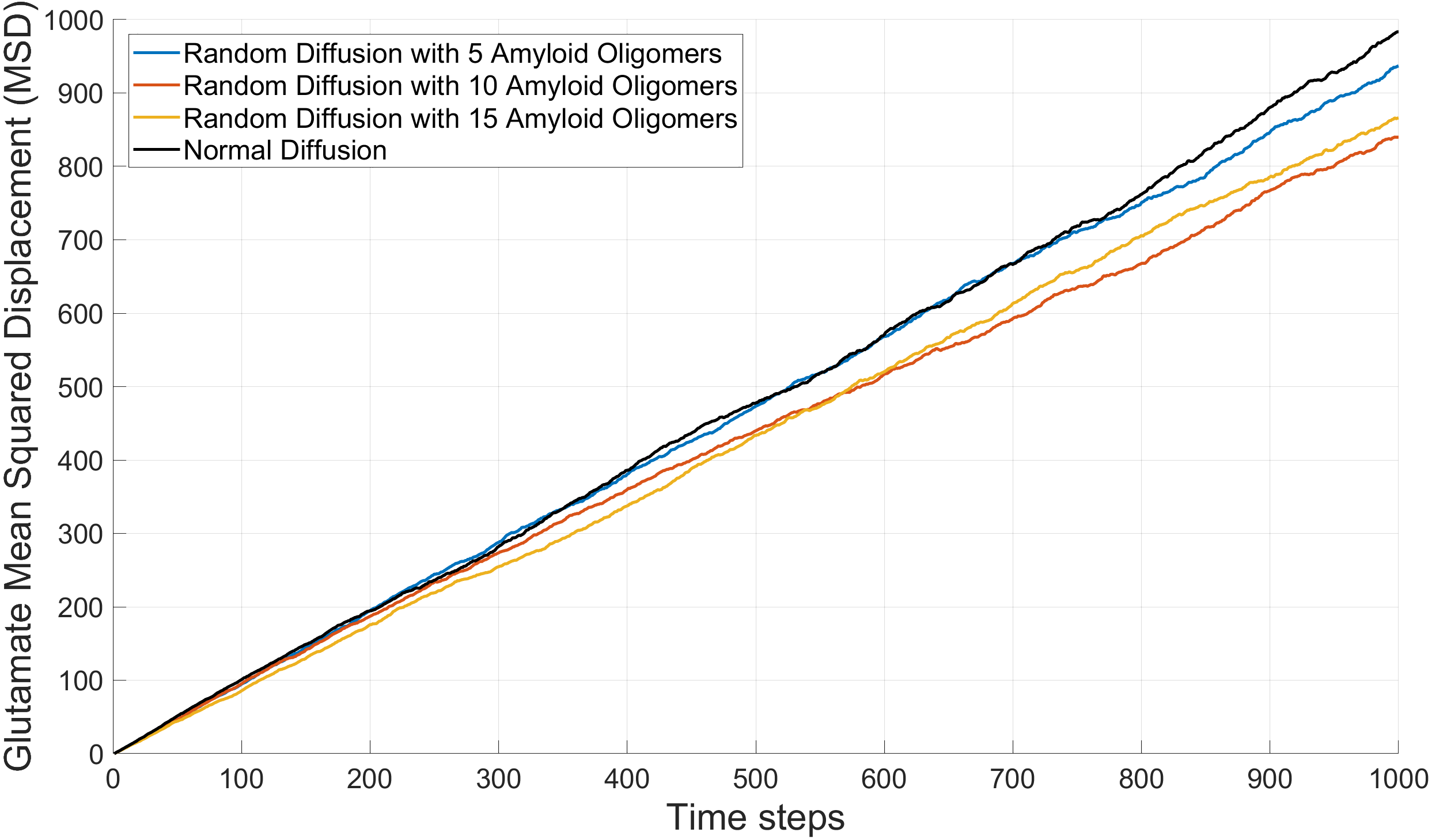}
    \caption{}
    \label{fig:MSD1}
  \end{subfigure}
  \hfill 
  \begin{subfigure}[b]{0.45\textwidth}
    \includegraphics[width=\textwidth]{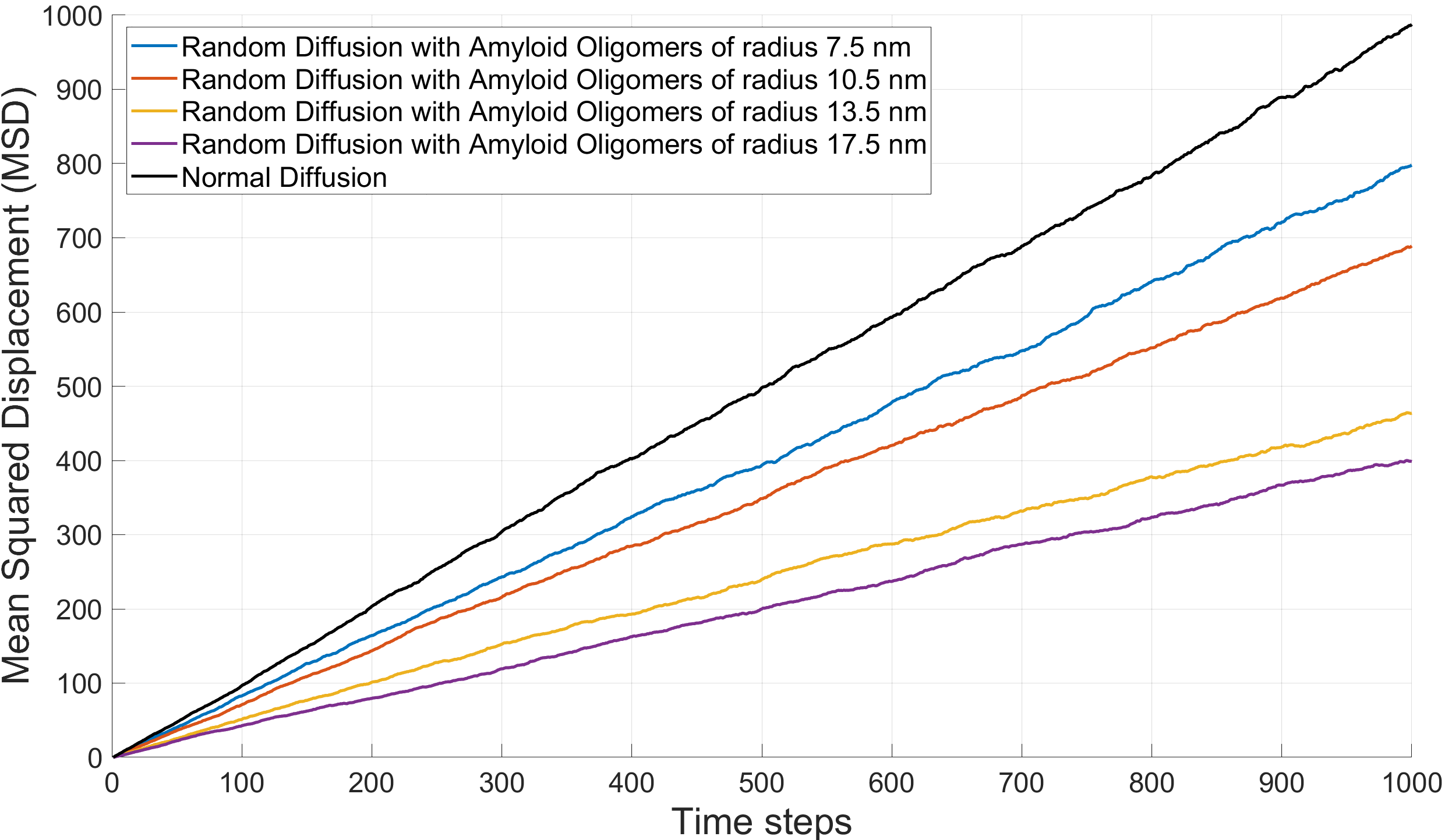}
    \caption{}
    \label{fig:MSD2}
  \end{subfigure}
 \caption{The effect of A$\beta$os on glutamate mean square displacement (MSD) within the synaptic cleft; (a) constant size of A$\beta$os with varying quantities, (b) constant quantity with varying sizes.}
  \label{Fig:MSDfull}
\end{figure}

Fig. \ref{Fig:MSDfull} illustrates the influence of A$\beta$os on MSD of glutamate molecules within the synaptic cleft, emphasizing the effects of both amyloid size and concentration. In Fig. \ref{fig:MSD1}, as the concentration of amyloid oligomers increases while their size remains constant, there is a marked reduction in the MSD of glutamate. This finding suggests that higher concentrations of A$\beta$os create more obstacles, thereby impeding glutamate diffusion. Conversely, Fig. \ref{fig:MSD2} presents the scenario where the concentration of A$\beta$os is constant, but their size varies according to the range of radii listed in Table \ref{tab:simulation_parameters}. A similar trend is observed: larger amyloid oligomers cause a significant decrease in glutamate MSD, indicating that larger obstacles more effectively obstruct glutamate movement. These findings collectively demonstrate that both the size and concentration of A$\beta$os are critical factors in disrupting glutamate diffusion, with larger and more numerous amyloid oligomers resulting in a more pronounced hindrance.

\begin{table}[h!]
\centering
\caption{Simulations Parameters}
\label{tab:simulation_parameters}
\begin{tabular}{|>{\centering\arraybackslash}m{4cm}|>{\centering\arraybackslash}m{4cm}|}
\hline
\textbf{Parameter} & \textbf{Value} \\
\hline
Coefficient of Restitution (e) & 0.6 \\
\hline
Collision Angles ($\theta$, $\phi$) & $30^\circ$ \\
\hline
Amyloid Radius & $7.5 < R< 17.5$ nm \\
\hline
Amyloid Mass & $>50$ kDa \\
\hline
Normal Diffusion Coefficient & $4.0 \times 10^{-3}$ $\mu$m$^2$/s \\
\hline
Glutamate Diffusion Coefficient & 300 $\mu$m$^2$/s \\
\hline
Initial Glutamate Concentration & 2000 molecules \\
\hline
Glutamate Radius & 0.5 Å \\
\hline
Glutamate Mass & 147.13 Daltons \\
\hline
Synaptic Cleft Legnth & 300 nm \\
\hline
Geometrical Configuration ($\lambda$) & 1,2,3 \\
\hline
Percolation Threshold ($\phi_c$) & 0.3, 0.5, 0.7 \\
\hline
\end{tabular}
\end{table}

\subsection{\texorpdfstring{A$\beta$o Movement}{Aβ Movement}}
In Section \ref{section:IV:C:2}, we explored how the kinematic equations for A$\beta$os can be employed to estimate the distance these particles travel post-collision with glutamate molecules. Considering the stochastic nature of A$\beta$o movement, which follows Brownian motion, we utilize the (\( v_{\text{rms}} \)) as a representative value for the relative velocity of both glutamate and A$\beta$os prior to collisions which are listed in Table \ref{table1}. These velocities are detailed in Table \ref{table1}.

\begin{figure}[ht]
\centering
\includegraphics[width=0.9\linewidth]{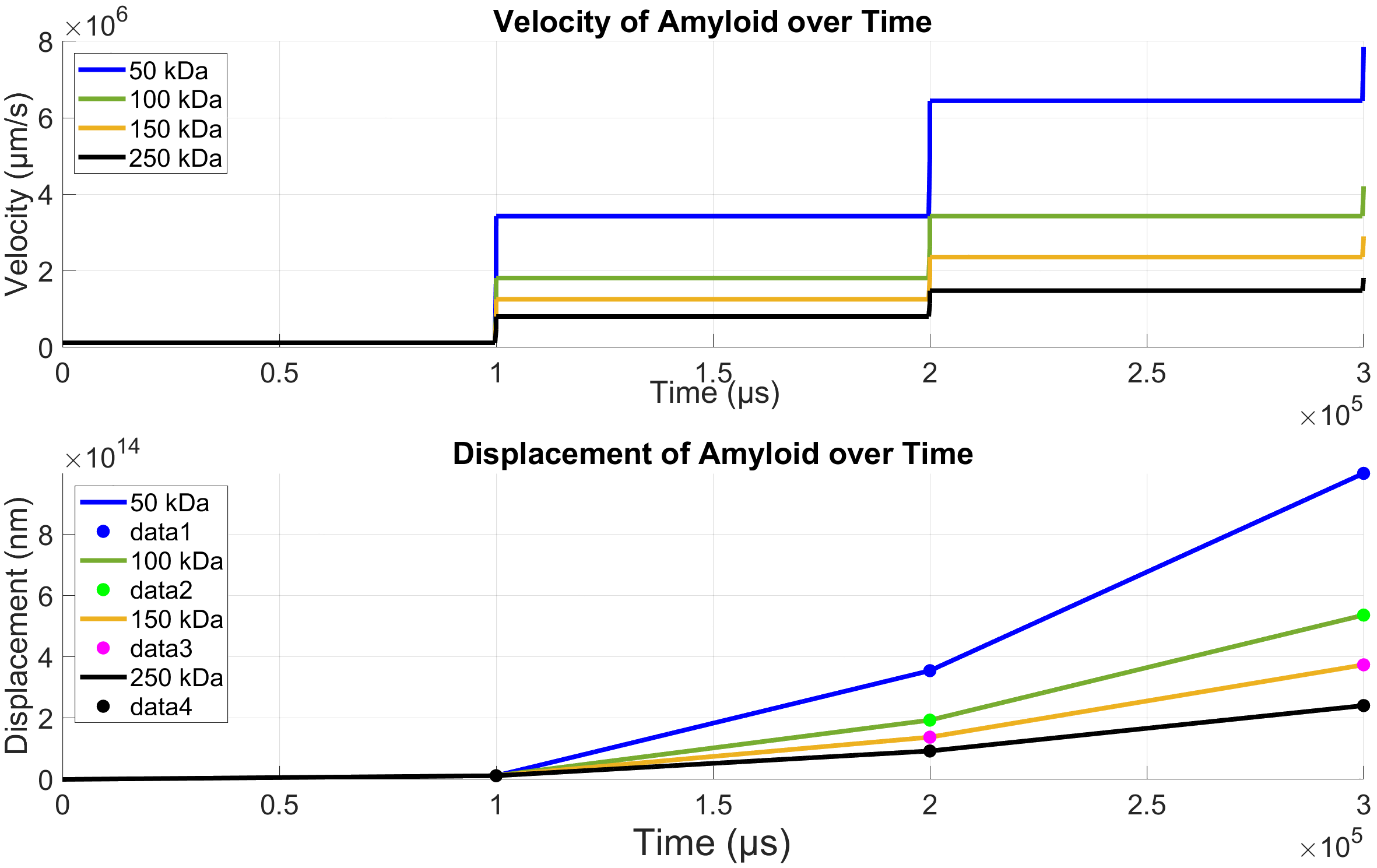}
\caption{Increasing A$\beta$ oligomer mass results in smaller increases in post-collision velocity and displacement.}
\label{fig:Colsize}
\end{figure}

Utilizing the equations for momentum (\ref{eq: momentum}) and elasticity (\ref{eq: elasticity}), Fig. \ref{fig:Colsize} illustrates the results of three consecutive collisions for A$\beta$os within the synaptic cleft. The simulation parameters used are detailed in Table \ref{tab:simulation_parameters}.

The graph shows a progressive increase in velocity with each collision, particularly evident in the first graph (Velocity of A$\beta$o over Time). The velocity steps up distinctly after each collision, indicating that the kinetic energy imparted during the collisions propels the A$\beta$os forward.

However, it is crucial to observe that the magnitude of these velocity increases—and consequently the displacement—is dependent on the size of the A$\beta$os. The second graph (Displacement of A$\beta$os over Time) shows that smaller oligomers (e.g., 50 kDa) exhibit much greater displacement over time compared to larger ones (e.g., 250 kDa). This is attributed to their lower mass, which translates to less inertia, allowing them to be more easily accelerated by the collisions.

Conversely, larger A$\beta$os, while still experiencing increases in velocity, show markedly smaller displacements. This is due to their greater mass and inertia, which makes them less responsive to the forces exerted during collisions.

These findings emphasize the significant role that both size and mass play in influencing the post-collision movement of A$\beta$os within the synaptic cleft. The analysis suggests that smaller A$\beta$os are more likely to travel further and potentially reach the postsynaptic membrane, potentially impacting synaptic function more than their larger counterparts.

\subsection{\texorpdfstring{A$\beta$o and NMDARs Binding}{Aβ binding}}
\begin{figure}[ht]
  \centering
  \includegraphics[width=0.5\textwidth]{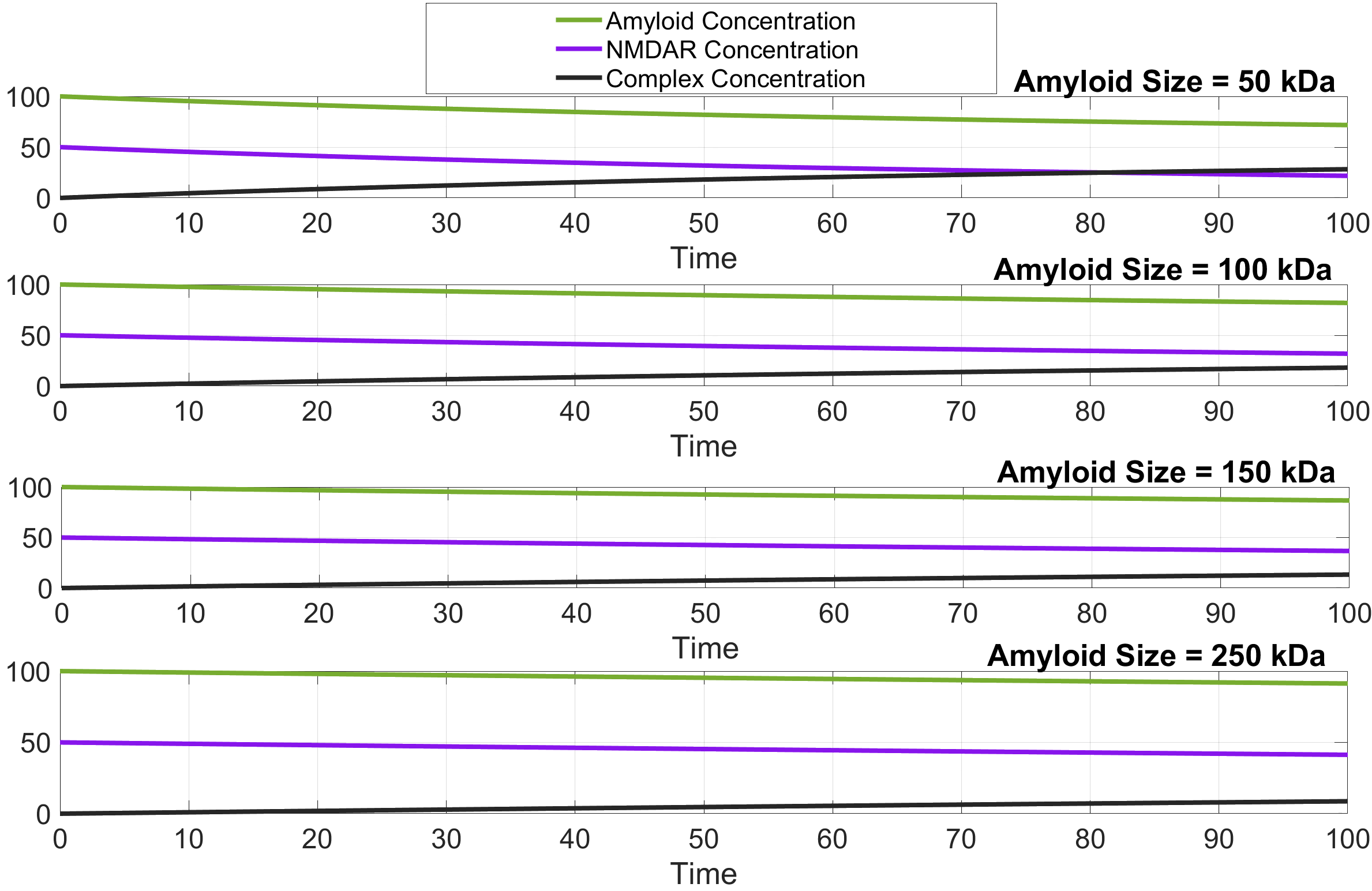}
  \caption{Kinetic model simulation for Amyloid=100, NMDARs= 50.}
  \label{fig: Bindingsize}
\end{figure}

In Section \ref{section:IV:C:3}, we propose that A$\beta$os are likely to bind to NMDARs as a result of their movement after collisions. Fig. \ref{fig: Bindingsize} presents four plots that depict the concentrations of A$\beta$os, NMDARs, and the Amyloid-NMDAR complex over time, with varying masses of A$\beta$os. The initial concentrations are assumed to be: A$\beta$os = 100 and NMDARs = 50.

The figure illustrates a clear trend: smaller A$\beta$ oligomers (50 kDa) exhibit a significantly higher propensity to bind to NMDARs compared to larger ones (e.g., 250 kDa). This trend is observed in the increasing concentration of the Amyloid-NMDAR complex as the mass of A$\beta$os decreases. In contrast, larger A$\beta$ oligomers show a relatively lower increase in complex concentration over the same time period.

This behavior can be attributed to the increased mobility of smaller A$\beta$ oligomers, which enhances their chances of encountering and binding to NMDARs. As these smaller A$\beta$os bind to NMDARs, they obstruct further receptor activation, leading to a weakening of LTP. This interference with normal synaptic function can disrupt subsequent signaling processes, potentially culminating in cellular damage and death \cite{huang2020toxicity}.

These findings are consistent with the observations in \cite{walsh2005certain}, which assert that LTP is primarily mediated by low-n oligomers (small oligomers), rather than A$\beta$ monomers or larger aggregates. The higher binding affinity of smaller A$\beta$os to NMDARs suggests that they play a more critical role in the disruption of synaptic signaling, which is a hallmark of neurodegenerative processes.

\subsection{SNR}
Through simulations using mathematical modeling for SNR in Sec. \ref{section:IV:D}, we assess the SNR under varying conditions, such as different concentrations and sizes of A$\beta$os. 

\begin{figure}[htbp]
    \centering
    \begin{subfigure}[b]{0.45\textwidth}
        \centering
        \includegraphics[width=\textwidth]{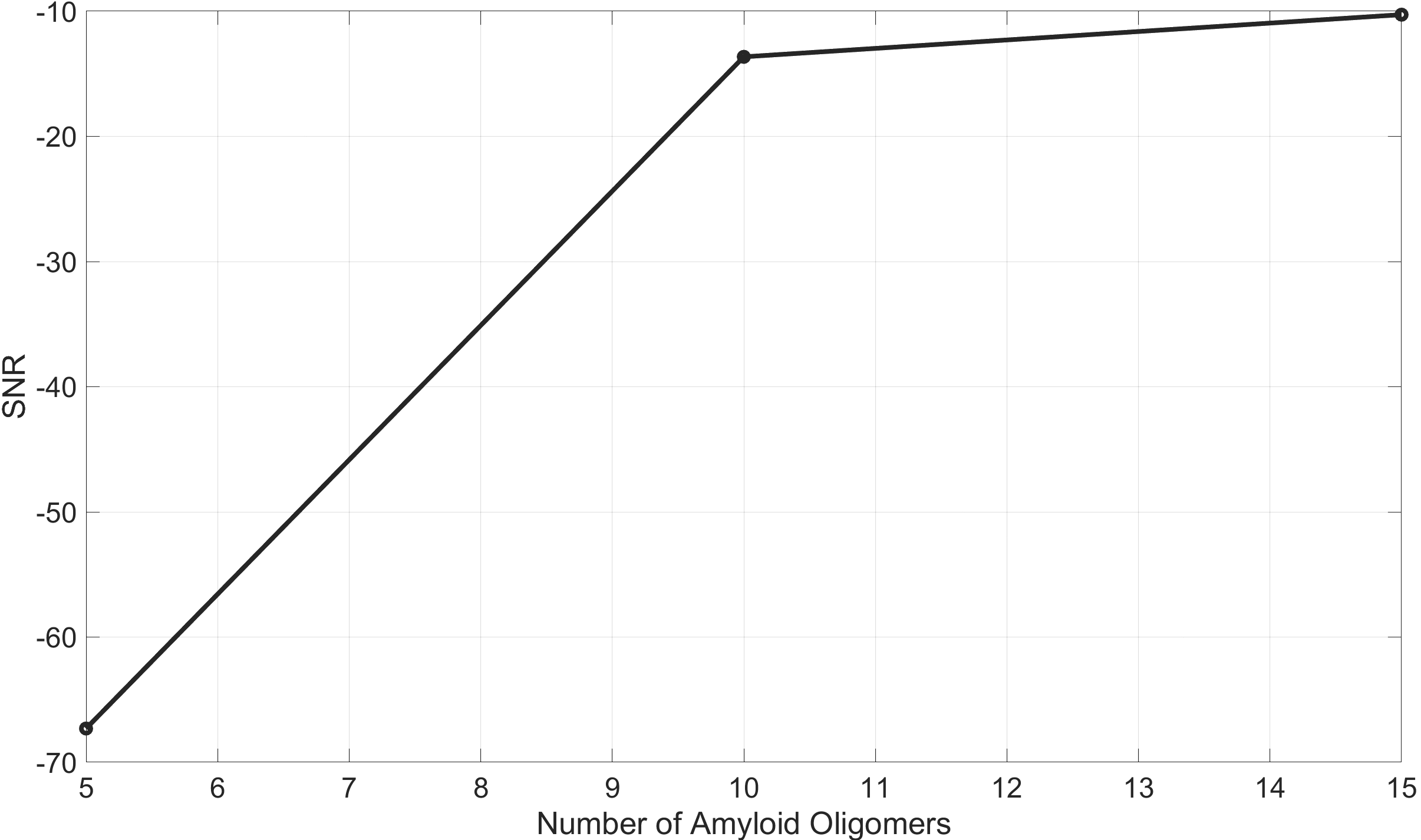}
        \caption{}
        \label{fig: SNRconcentration}
    \end{subfigure}
    \hfill
    \begin{subfigure}[b]{0.45\textwidth}
        \centering
        \includegraphics[width=\textwidth]{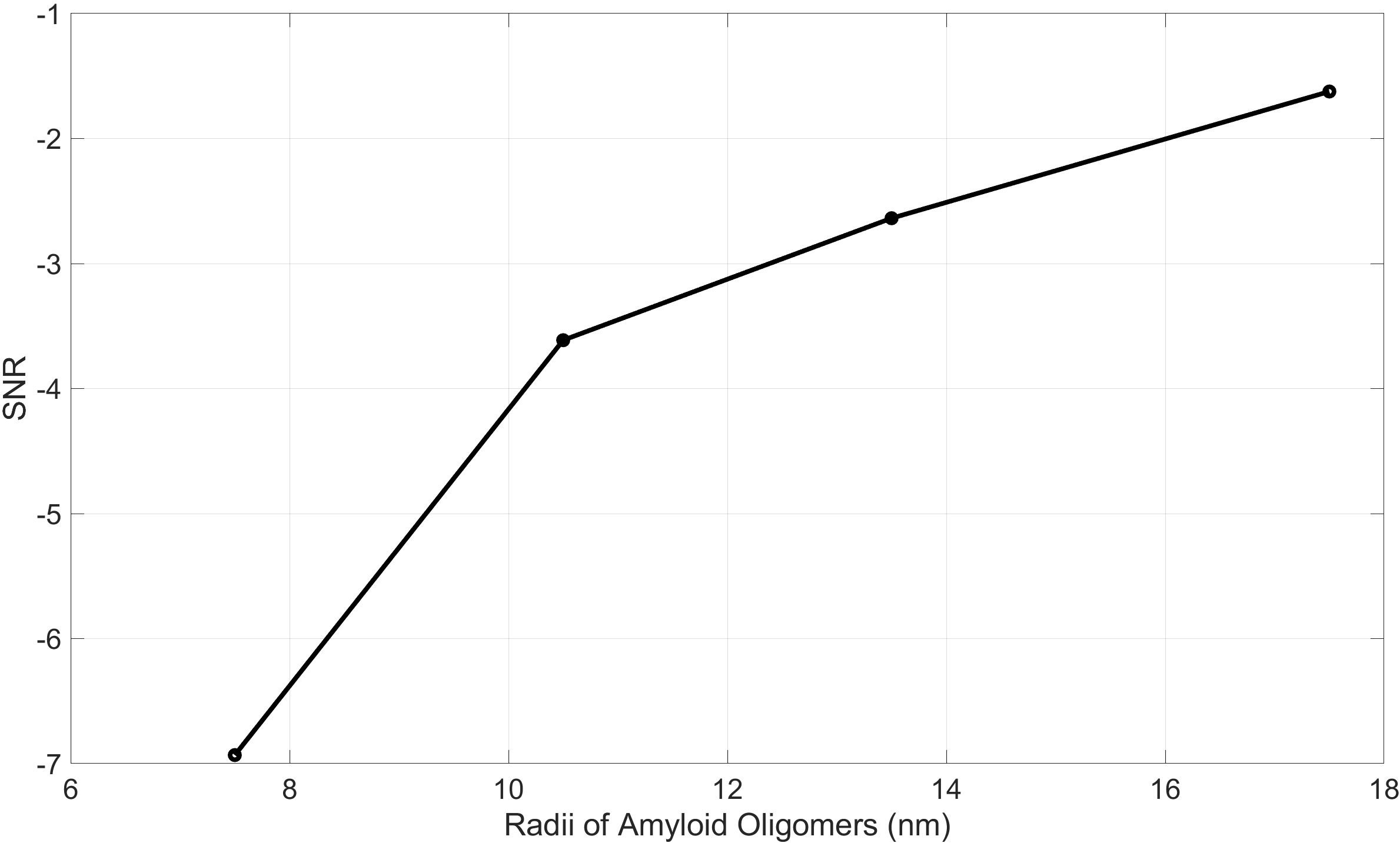}
        \caption{}
        \label{fig: SNRsize}
    \end{subfigure}
    \caption{SNR vs. Radii of Amyloid Oligomers (a), and SNR vs. Number of Amyloid Oligomers (b)}
    \label{fig:SNR}
\end{figure}

The impact of A$\beta$os on the synaptic cleft as a molecular communication channel was evaluated by analyzing the SNR, which serves as a critical metric for understanding the level of noise introduced into the communication channel by these obstacles. The SNR provides a quantitative measure of the channel's efficiency in transmitting signals (in this case, molecular signals such as glutamate) in the presence of noise-inducing factors A$\beta$os. Two primary variables were considered in : the concentration (number) of A$\beta$os and their size (radius).
As shown in Fig. \ref{fig: SNRconcentration}, the SNR decreases significantly with an increasing number of A$\beta$os, indicating a rise in channel noise. At a low concentration of five A$\beta$os, the channel exhibits a high level of noise, as evidenced by the highly negative SNR of approximately -70. This suggests that even a small number of obstacles can severely degrade the quality of molecular communication within the synaptic cleft. As the number of A$\beta$os increases to ten and fifteen, the SNR improves (becomes less negative), reflecting a non-linear relationship between obstacle concentration and channel noise. The saturation effect observed implies that while additional A$\beta$os continue to introduce noise, their relative impact diminishes as the channel becomes increasingly congested. This trend indicates that the channel noise increases sharply at lower concentrations but stabilizes at higher concentrations, possibly due to the overlapping influence of multiple obstacles.

The analysis of the impact of A$\beta$os size on channel noise, as illustrated in Fig. \ref{fig: SNRsize}, reveals that smaller oligomers (with a radius of 7.5 nm) introduce greater noise into the molecular communication channel, resulting in a lower SNR of approximately -7. As the size of the oligomers increases, the SNR improves, indicating a reduction in channel noise. Larger oligomers, although still obstructive, appear to create a more stable and less disruptive communication environment. This finding suggests that larger oligomers may cause more predictable and consistent noise, leading to a less severe impact on the overall communication process compared to smaller oligomers, which cause more random and disruptive noise patterns.
Finally, the study highlights the significant role that both the concentration and size of A$\beta$os play in determining the noise levels within the synaptic cleft as a molecular communication channel. While higher concentrations of A$\beta$os substantially increase channel noise, the impact becomes less pronounced at higher concentrations, likely due to saturation effects. On the other hand, smaller amyloid A$\beta$os generate more severe noise compared to larger ones, underscoring the importance of obstacle size in the modulation of channel noise. These findings provide crucial insights into the challenges of maintaining effective molecular communication in biological systems, particularly in pathological conditions such as AD, where A$\beta$os are prevalent and contribute to impaired communication by introducing significant noise into the synaptic channel.

Overall, the results of our study are align with the experiments in vitro and in vivo analysis of \cite{shankar2007natural, cleary2005natural} which have demonstrated that cell-derived low-n (small oligomers) A$\beta$os can trigger hippocampal synapse loss and may be important effectors of synaptic dysfunction in AD.

\section{Conclusion and Future Directions}
\label{section:VI}

Our study utilized stochastic modeling to investigate the interactions between A$\beta$os and glutamate within the synaptic cleft, with a particular emphasis on how the size and mass of A$\beta$os influence glutamate diffusion, concentration, and displacement. The findings demonstrate that the size of A$\beta$os is a pivotal factor in determining their impact on synaptic function. Smaller oligomers, due to their higher mobility, are more likely to bind with NMDARs, highlighting them as potential focal points for therapeutic interventions. In contrast, larger oligomers, which exhibit reduced mobility, are less likely to bind with postsynaptic membrane receptors but significantly disrupt glutamate dynamics by reducing its velocity following collisions. These insights emphasize the need to consider the diverse physical properties of A$\beta$os in the development of targeted therapies aimed at mitigating synaptic disruptions in AD.

The findings of this study invite further exploration into several promising areas. An in-depth analysis of the collision dynamics between A$\beta$os and glutamate molecules could elucidate the synaptic disruptions observed in AD. Specifically, understanding how the size of A$\beta$os affects their interaction with glutamate may unveil new therapeutic targets aimed at mitigating synaptic toxicity. Furthermore, strategies to shrink or stabilize larger oligomers should be investigated to reduce their obstructive impact on glutamate diffusion and, consequently, their contribution to synaptic dysfunction.

Advancements in imaging technologies are crucial for progressing this research field. Developing high-resolution imaging systems capable of accurately measuring A$\beta$os size in real-time would significantly enhance our ability to monitor the progression of AD and evaluate the effectiveness of therapeutic interventions. Additionally, the creation of a real-time analytical system to dynamically observe interactions within the synaptic cleft could provide invaluable insights into the mechanistic processes disrupting synaptic function.

On the therapeutic front, targeted interventions should focus on manipulating NMDAR interactions with smaller oligomers. Enhancing these interactions could potentially offset the adverse effects caused by larger oligomers, thus preserving synaptic function. This approach warrants the development of pharmacological agents specifically tailored to the characteristics of A$\beta$os, such as their size and mobility. Such specificity could lead to more effective treatments that directly address the pathological features of AD.


%

\bibliographystyle{IEEEtran}
\bibliography{IEEEabrv, Reference}

\begin{thebibliography}{10}
\providecommand{\url}[1]{#1}
\csname url@samestyle\endcsname
\providecommand{\newblock}{\relax}
\providecommand{\bibinfo}[2]{#2}
\providecommand{\BIBentrySTDinterwordspacing}{\spaceskip=0pt\relax}
\providecommand{\BIBentryALTinterwordstretchfactor}{4}
\providecommand{\BIBentryALTinterwordspacing}{\spaceskip=\fontdimen2\font plus
\BIBentryALTinterwordstretchfactor\fontdimen3\font minus \fontdimen4\font\relax}
\providecommand{\BIBforeignlanguage}[2]{{%
\expandafter\ifx\csname l@#1\endcsname\relax
\typeout{** WARNING: IEEEtran.bst: No hyphenation pattern has been}%
\typeout{** loaded for the language `#1'. Using the pattern for}%
\typeout{** the default language instead.}%
\else
\language=\csname l@#1\endcsname
\fi
#2}}
\providecommand{\BIBdecl}{\relax}
\BIBdecl

\bibitem{diogo2022early}
V.~S. Diogo, H.~A. Ferreira, D.~Prata, and A.~D.~N. Initiative, ``Early diagnosis of alzheimer’s disease using machine learning: a multi-diagnostic, generalizable approach,'' \emph{Alzheimer's Research \& Therapy}, vol.~14, no.~1, p. 107, 2022.

\bibitem{mebane20092009}
I.~Mebane-Sims, ``2009 alzheimer's disease facts and figures.'' \emph{Alzheimer's \& Dementia: The Journal of the Alzheimer's Association}, 2009.

\bibitem{tolar2021neurotoxic}
M.~Tolar, J.~Hey, A.~Power, and S.~Abushakra, ``Neurotoxic soluble amyloid oligomers drive alzheimer’s pathogenesis and represent a clinically validated target for slowing disease progression,'' \emph{International journal of molecular sciences}, vol.~22, no.~12, p. 6355, 2021.

\bibitem{matthews2021riluzole}
D.~C. Matthews, X.~Mao, K.~Dowd, D.~Tsakanikas, C.~S. Jiang, C.~Meuser, R.~D. Andrews, A.~S. Lukic, J.~Lee, N.~Hampilos \emph{et~al.}, ``Riluzole, a glutamate modulator, slows cerebral glucose metabolism decline in patients with alzheimer’s disease,'' \emph{Brain}, vol. 144, no.~12, pp. 3742--3755, 2021.

\bibitem{morrison2002selective}
J.~H. Morrison and P.~R. Hof, ``Selective vulnerability of corticocortical and hippocampal circuits in aging and alzheimer's disease,'' \emph{Progress in brain research}, vol. 136, pp. 467--486, 2002.

\bibitem{hof1990quantitative}
P.~R. Hof and J.~H. Morrison, ``Quantitative analysis of a vulnerable subset of pyramidal neurons in alzheimer's disease: Ii. primary and secondary visual cortex,'' \emph{Journal of Comparative Neurology}, vol. 301, no.~1, pp. 55--64, 1990.

\bibitem{li2009soluble}
S.~Li, S.~Hong, N.~E. Shepardson, D.~M. Walsh, G.~M. Shankar, and D.~Selkoe, ``Soluble oligomers of amyloid $\beta$ protein facilitate hippocampal long-term depression by disrupting neuronal glutamate uptake,'' \emph{Neuron}, vol.~62, no.~6, pp. 788--801, 2009.

\bibitem{kamenetz2003app}
F.~Kamenetz, T.~Tomita, H.~Hsieh, G.~Seabrook, D.~Borchelt, T.~Iwatsubo, S.~Sisodia, and R.~Malinow, ``App processing and synaptic function,'' \emph{Neuron}, vol.~37, no.~6, pp. 925--937, 2003.

\bibitem{snyder2005regulation}
E.~M. Snyder, Y.~Nong, C.~G. Almeida, S.~Paul, T.~Moran, E.~Y. Choi, A.~C. Nairn, M.~W. Salter, P.~J. Lombroso, G.~K. Gouras \emph{et~al.}, ``Regulation of nmda receptor trafficking by amyloid-$\beta$,'' \emph{Nature neuroscience}, vol.~8, no.~8, pp. 1051--1058, 2005.

\bibitem{mcentee1993glutamate}
W.~J. McEntee and T.~H. Crook, ``Glutamate: its role in learning, memory, and the aging brain,'' \emph{Psychopharmacology}, vol. 111, no.~4, pp. 391--401, 1993.

\bibitem{schousboe1981transport}
A.~Schousboe, ``Transport and metabolism of glutamate and gaba in neurons and glial cells,'' \emph{International review of neurobiology}, vol.~22, pp. 1--45, 1981.

\bibitem{rubenstein2003model}
J.~Rubenstein and M.~M. Merzenich, ``Model of autism: increased ratio of excitation/inhibition in key neural systems,'' \emph{Genes, Brain and Behavior}, vol.~2, no.~5, pp. 255--267, 2003.

\bibitem{wu2012role}
K.~Wu, G.~L. Hanna, D.~R. Rosenberg, and P.~D. Arnold, ``The role of glutamate signaling in the pathogenesis and treatment of obsessive--compulsive disorder,'' \emph{Pharmacology Biochemistry and Behavior}, vol. 100, no.~4, pp. 726--735, 2012.

\bibitem{gao2015common}
R.~Gao and P.~Penzes, ``Common mechanisms of excitatory and inhibitory imbalance in schizophrenia and autism spectrum disorders,'' \emph{Current molecular medicine}, vol.~15, no.~2, pp. 146--167, 2015.

\bibitem{huang2020toxicity}
Y.-r. Huang and R.-t. Liu, ``The toxicity and polymorphism of $\beta$-amyloid oligomers,'' \emph{International journal of molecular sciences}, vol.~21, no.~12, p. 4477, 2020.

\bibitem{mroczko2018amyloid}
B.~Mroczko, M.~Groblewska, A.~Litman-Zawadzka, J.~Kornhuber, and P.~Lewczuk, ``Amyloid $\beta$ oligomers (a$\beta$os) in alzheimer’s disease,'' \emph{Journal of Neural Transmission}, vol. 125, pp. 177--191, 2018.

\bibitem{taniguchi2022amyloid}
K.~Taniguchi, F.~Yamamoto, A.~Amano, A.~Tamaoka, N.~Sanjo, T.~Yokota, F.~Kametani, and W.~Araki, ``Amyloid-$\beta$ oligomers interact with nmda receptors containing glun2b subunits and metabotropic glutamate receptor 1 in primary cortical neurons: Relevance to the synapse pathology of alzheimer’s disease,'' \emph{Neuroscience Research}, vol. 180, pp. 90--98, 2022.

\bibitem{owen2019effects}
M.~C. Owen, D.~Gnutt, M.~Gao, S.~K. W{\"a}rml{\"a}nder, J.~Jarvet, A.~Gr{\"a}slund, R.~Winter, S.~Ebbinghaus, and B.~Strodel, ``Effects of in vivo conditions on amyloid aggregation,'' \emph{Chemical Society Reviews}, vol.~48, no.~14, pp. 3946--3996, 2019.

\bibitem{lucˇic2005morphological}
V.~Lucˇic, T.~Yang, G.~Schweikert, F.~F{\"o}rster, and W.~Baumeister, ``Morphological characterization of molecular complexes present in the synaptic cleft,'' \emph{Structure}, vol.~13, no.~3, pp. 423--434, 2005.

\bibitem{meldrum2000glutamate}
B.~S. Meldrum, ``Glutamate as a neurotransmitter in the brain: review of physiology and pathology,'' \emph{The Journal of nutrition}, vol. 130, no.~4, pp. 1007S--1015S, 2000.

\bibitem{cijsouw2018mapping}
T.~Cijsouw, A.~M. Ramsey, T.~T. Lam, B.~E. Carbone, T.~A. Blanpied, and T.~Biederer, ``Mapping the proteome of the synaptic cleft through proximity labeling reveals new cleft proteins,'' \emph{Proteomes}, vol.~6, no.~4, p.~48, 2018.

\bibitem{zheng2017nanoscale}
K.~Zheng, T.~P. Jensen, L.~P. Savtchenko, J.~A. Levitt, K.~Suhling, and D.~A. Rusakov, ``Nanoscale diffusion in the synaptic cleft and beyond measured with time-resolved fluorescence anisotropy imaging,'' \emph{Scientific reports}, vol.~7, no.~1, p. 42022, 2017.

\bibitem{kulish2019modeling}
O.~Kulish and A.~Vasilev, ``Modeling the nerve impulse transmission in a synaptic cleft,'' \emph{Journal of Physical Research}, no. 23, Number 1, pp. 1801--1, 2019.

\bibitem{akan2016fundamentals}
O.~B. Akan, H.~Ramezani, T.~Khan, N.~A. Abbasi, and M.~Kuscu, ``Fundamentals of molecular information and communication science,'' \emph{Proceedings of the IEEE}, vol. 105, no.~2, pp. 306--318, 2016.

\bibitem{balevi2013physical}
E.~Balevi and O.~B. Akan, ``A physical channel model for nanoscale neuro-spike communications,'' \emph{IEEE Transactions on Communications}, vol.~61, no.~3, pp. 1178--1187, 2013.

\bibitem{malak2013communication}
D.~Malak and O.~B. Akan, ``A communication theoretical analysis of synaptic multiple-access channel in hippocampal-cortical neurons,'' \emph{IEEE Transactions on communications}, vol.~61, no.~6, pp. 2457--2467, 2013.

\bibitem{ramezani2017information}
H.~Ramezani and O.~B. Akan, ``Information capacity of vesicle release in neuro-spike communication,'' \emph{IEEE Communications Letters}, vol.~22, no.~1, pp. 41--44, 2017.

\bibitem{tu2014oligomeric}
S.~Tu, S.-i. Okamoto, S.~A. Lipton, and H.~Xu, ``Oligomeric a$\beta$-induced synaptic dysfunction in alzheimer’s disease,'' \emph{Molecular neurodegeneration}, vol.~9, pp. 1--12, 2014.

\bibitem{hayashi2018structure}
M.~K. Hayashi, ``Structure-function relationship of transporters in the glutamate--glutamine cycle of the central nervous system,'' \emph{International journal of molecular sciences}, vol.~19, no.~4, p. 1177, 2018.

\bibitem{clements1992time}
J.~D. Clements, R.~A. Lester, G.~Tong, C.~E. Jahr, and G.~L. Westbrook, ``The time course of glutamate in the synaptic cleft,'' \emph{Science}, vol. 258, no. 5087, pp. 1498--1501, 1992.

\bibitem{kayed2013molecular}
R.~Kayed and C.~A. Lasagna-Reeves, ``Molecular mechanisms of amyloid oligomers toxicity,'' \emph{Journal of Alzheimer's Disease}, vol.~33, no.~s1, pp. S67--S78, 2013.

\bibitem{rinauro2024misfolded}
D.~J. Rinauro, F.~Chiti, M.~Vendruscolo, and R.~Limbocker, ``Misfolded protein oligomers: mechanisms of formation, cytotoxic effects, and pharmacological approaches against protein misfolding diseases,'' \emph{Molecular Neurodegeneration}, vol.~19, no.~1, pp. 1--32, 2024.

\bibitem{du2022mitochondrial}
F.~Du, Q.~Yu, N.~M. Kanaan, and S.~S. Yan, ``Mitochondrial oxidative stress contributes to the pathological aggregation and accumulation of tau oligomers in alzheimer’s disease,'' \emph{Human Molecular Genetics}, vol.~31, no.~15, pp. 2498--2507, 2022.

\bibitem{kayed2007fibril}
R.~Kayed, E.~Head, F.~Sarsoza, T.~Saing, C.~W. Cotman, M.~Necula, L.~Margol, J.~Wu, L.~Breydo, J.~L. Thompson \emph{et~al.}, ``Fibril specific, conformation dependent antibodies recognize a generic epitope common to amyloid fibrils and fibrillar oligomers that is absent in prefibrillar oligomers,'' \emph{Molecular neurodegeneration}, vol.~2, pp. 1--11, 2007.

\bibitem{calamai2011single}
M.~Calamai and F.~S. Pavone, ``Single molecule tracking analysis reveals that the surface mobility of amyloid oligomers is driven by their conformational structure,'' \emph{Journal of the American Chemical Society}, vol. 133, no.~31, pp. 12\,001--12\,008, 2011.

\bibitem{studysmarter_sde}
{StudySmarter}, ``Stochastic differential equations: Overview,'' \url{https://www.studysmarter.co.uk/explanations/advanced-mathematics/stochastic-differential-equations/}, accessed: 2024-07-31.

\bibitem{riahi2019identifying}
M.~Riahi, I.~Qattan, J.~Hassan, and D.~Homouz, ``Identifying short-and long-time modes of the mean-square displacement: An improved nonlinear fitting approach,'' \emph{AIP Advances}, vol.~9, no.~5, 2019.

\bibitem{bannai2006imaging}
H.~Bannai, S.~L{\'e}vi, C.~Schweizer, M.~Dahan, and A.~Triller, ``Imaging the lateral diffusion of membrane molecules with quantum dots,'' \emph{Nature protocols}, vol.~1, no.~6, pp. 2628--2634, 2006.

\bibitem{ehrensperger2007multiple}
M.-V. Ehrensperger, C.~Hanus, C.~Vannier, A.~Triller, and M.~Dahan, ``Multiple association states between glycine receptors and gephyrin identified by spt analysis,'' \emph{Biophysical journal}, vol.~92, no.~10, pp. 3706--3718, 2007.

\bibitem{chimon2007evidence}
S.~Chimon, M.~A. Shaibat, C.~R. Jones, D.~C. Calero, B.~Aizezi, and Y.~Ishii, ``Evidence of fibril-like $\beta$-sheet structures in a neurotoxic amyloid intermediate of alzheimer's $\beta$-amyloid,'' \emph{Nature structural \& molecular biology}, vol.~14, no.~12, pp. 1157--1164, 2007.

\bibitem{lacor2007abeta}
P.~N. Lacor, M.~C. Buniel, P.~W. Furlow, A.~S. Clemente, P.~T. Velasco, M.~Wood, K.~L. Viola, and W.~L. Klein, ``A$\beta$ oligomer-induced aberrations in synapse composition, shape, and density provide a molecular basis for loss of connectivity in alzheimer's disease,'' \emph{Journal of Neuroscience}, vol.~27, no.~4, pp. 796--807, 2007.

\bibitem{velasco2012synapse}
P.~T. Velasco, M.~C. Heffern, A.~Sebollela, I.~A. Popova, P.~N. Lacor, K.~B. Lee, X.~Sun, B.~N. Tiano, K.~L. Viola, A.~L. Eckermann \emph{et~al.}, ``Synapse-binding subpopulations of a$\beta$ oligomers sensitive to peptide assembly blockers and scfv antibodies,'' \emph{ACS chemical neuroscience}, vol.~3, no.~11, pp. 972--981, 2012.

\bibitem{li2022computational}
X.~Li, G.~H{\'e}mond, A.~G. Godin, and N.~Doyon, ``Computational modeling of trans-synaptic nanocolumns, a modulator of synaptic transmission,'' \emph{Frontiers in Computational Neuroscience}, vol.~16, p. 969119, 2022.

\bibitem{wathey1979numerical}
J.~C. Wathey, M.~M. Nass, and H.~A. Lester, ``Numerical reconstruction of the quantal event at nicotinic synapses,'' \emph{Biophysical Journal}, vol.~27, no.~1, pp. 145--164, 1979.

\bibitem{vilaseca2011new}
E.~Vilaseca, A.~Isvoran, S.~Madurga, I.~Pastor, J.~L. Garc{\'e}s, and F.~Mas, ``New insights into diffusion in 3d crowded media by monte carlo simulations: effect of size, mobility and spatial distribution of obstacles,'' \emph{Physical Chemistry Chemical Physics}, vol.~13, no.~16, pp. 7396--7407, 2011.

\bibitem{meyer2011particle}
C.~Meyer and D.~Deglon, ``Particle collision modeling--a review,'' \emph{Minerals Engineering}, vol.~24, no.~8, pp. 719--730, 2011.

\bibitem{burko2019two}
L.~M. Burko, ``Two-dimensional collisions and conservation of momentum,'' \emph{The Physics Teacher}, vol.~57, no.~7, pp. 487--489, 2019.

\bibitem{novak2011diffusion}
I.~L. Novak, F.~Gao, P.~Kraikivski, and B.~M. Slepchenko, ``Diffusion amid random overlapping obstacles: Similarities, invariants, approximations,'' \emph{The Journal of chemical physics}, vol. 134, no.~15, 2011.

\bibitem{liu2019role}
J.~Liu, L.~Chang, Y.~Song, H.~Li, and Y.~Wu, ``The role of nmda receptors in alzheimer’s disease,'' \emph{Frontiers in neuroscience}, vol.~13, p.~43, 2019.

\bibitem{paoletti2013nmda}
P.~Paoletti, C.~Bellone, and Q.~Zhou, ``Nmda receptor subunit diversity: impact on receptor properties, synaptic plasticity and disease,'' \emph{Nature Reviews Neuroscience}, vol.~14, no.~6, pp. 383--400, 2013.

\bibitem{kodis2018n}
E.~J. Kodis, S.~Choi, E.~Swanson, G.~Ferreira, and G.~S. Bloom, ``N-methyl-d-aspartate receptor--mediated calcium influx connects amyloid-$\beta$ oligomers to ectopic neuronal cell cycle reentry in alzheimer's disease,'' \emph{Alzheimer's \& Dementia}, vol.~14, no.~10, pp. 1302--1312, 2018.

\bibitem{macdermott1986nmda}
A.~B. MacDermott, M.~L. Mayer, G.~L. Westbrook, S.~J. Smith, and J.~L. Barker, ``Nmda-receptor activation increases cytoplasmic calcium concentration in cultured spinal cord neurones,'' \emph{Nature}, vol. 321, no. 6069, pp. 519--522, 1986.

\bibitem{muller2009both}
T.~M{\"u}ller, D.~Albrecht, and C.~Gebhardt, ``Both nr2a and nr2b subunits of the nmda receptor are critical for long-term potentiation and long-term depression in the lateral amygdala of horizontal slices of adult mice,'' \emph{Learning \& Memory}, vol.~16, no.~6, pp. 395--405, 2009.

\bibitem{villmann2007hypes}
C.~Villmann and C.-M. Becker, ``On the hypes and falls in neuroprotection: targeting the nmda receptor,'' \emph{The Neuroscientist}, vol.~13, no.~6, pp. 594--615, 2007.

\bibitem{benarroch2011nmda}
E.~E. Benarroch, ``Nmda receptors: recent insights and clinical correlations,'' \emph{Neurology}, vol.~76, no.~20, pp. 1750--1757, 2011.

\bibitem{wenk2006neuropathologic}
G.~L. Wenk, ``Neuropathologic changes in alzheimer's disease: potential targets for treatment,'' \emph{Journal of Clinical Psychiatry}, vol.~67, p.~3, 2006.

\bibitem{malinow2012new}
R.~Malinow, ``New developments on the role of nmda receptors in alzheimer's disease,'' \emph{Current opinion in neurobiology}, vol.~22, no.~3, pp. 559--563, 2012.

\bibitem{verges2011opposing}
D.~K. Verges, J.~L. Restivo, W.~D. Goebel, D.~M. Holtzman, and J.~R. Cirrito, ``Opposing synaptic regulation of amyloid-$\beta$ metabolism by nmda receptors in vivo,'' \emph{Journal of Neuroscience}, vol.~31, no.~31, pp. 11\,328--11\,337, 2011.

\bibitem{resat2009kinetic}
H.~Resat, L.~Petzold, and M.~F. Pettigrew, ``Kinetic modeling of biological systems,'' \emph{Computational systems biology}, pp. 311--335, 2009.

\bibitem{borges2024cell}
L.~F. Borges, M.~T. Barros, and M.~Nogueira, ``Cell signaling error control for reliable molecular communications,'' \emph{Frontiers in Communications and Networks}, vol.~5, p. 1332379, 2024.

\bibitem{walsh2005certain}
D.~M. Walsh, M.~Townsend, M.~B. Podlisny, G.~M. Shankar, J.~V. Fadeeva, O.~El~Agnaf, D.~M. Hartley, and D.~J. Selkoe, ``Certain inhibitors of synthetic amyloid $\beta$-peptide (a$\beta$) fibrillogenesis block oligomerization of natural a$\beta$ and thereby rescue long-term potentiation,'' \emph{Journal of Neuroscience}, vol.~25, no.~10, pp. 2455--2462, 2005.

\bibitem{shankar2007natural}
G.~M. Shankar, B.~L. Bloodgood, M.~Townsend, D.~M. Walsh, D.~J. Selkoe, and B.~L. Sabatini, ``Natural oligomers of the alzheimer amyloid-$\beta$ protein induce reversible synapse loss by modulating an nmda-type glutamate receptor-dependent signaling pathway,'' \emph{Journal of Neuroscience}, vol.~27, no.~11, pp. 2866--2875, 2007.

\bibitem{cleary2005natural}
J.~P. Cleary, D.~M. Walsh, J.~J. Hofmeister, G.~M. Shankar, M.~A. Kuskowski, D.~J. Selkoe, and K.~H. Ashe, ``Natural oligomers of the amyloid-$\beta$ protein specifically disrupt cognitive function,'' \emph{Nature neuroscience}, vol.~8, no.~1, pp. 79--84, 2005.

\end{thebibliography}

%




\begin{IEEEbiography}[{\includegraphics[width=1in,height=1.25in,clip,keepaspectratio]{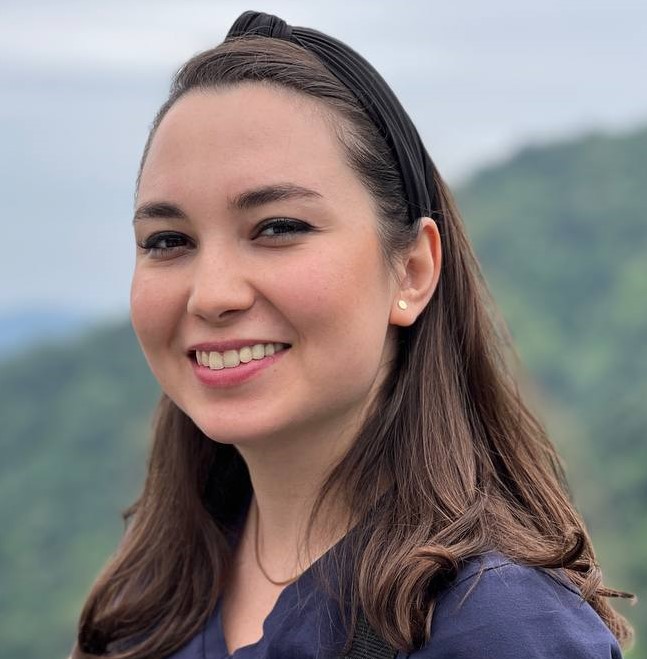}}]{Nayereh FallahBagheri}(PhD. Student Member, IEEE)
received her BSc in Electrical Engineering from Semnan State University, Semnan, Iran, in 2018. She is currently working as a research assistant at the Next-generation and Wireless Communications Laboratory and pursuing her Ph.D. degree in Electrical and Electronics Engineering at Koc University, under the supervision of Prof. Dr. Özgür Barış Akan. Her
 research interests include molecular communications, Intrabody Nanonetworks and the Internet of Everything.
\end{IEEEbiography}
\begin{IEEEbiography}[{\includegraphics[width=1in,height=1.25in,clip,keepaspectratio]{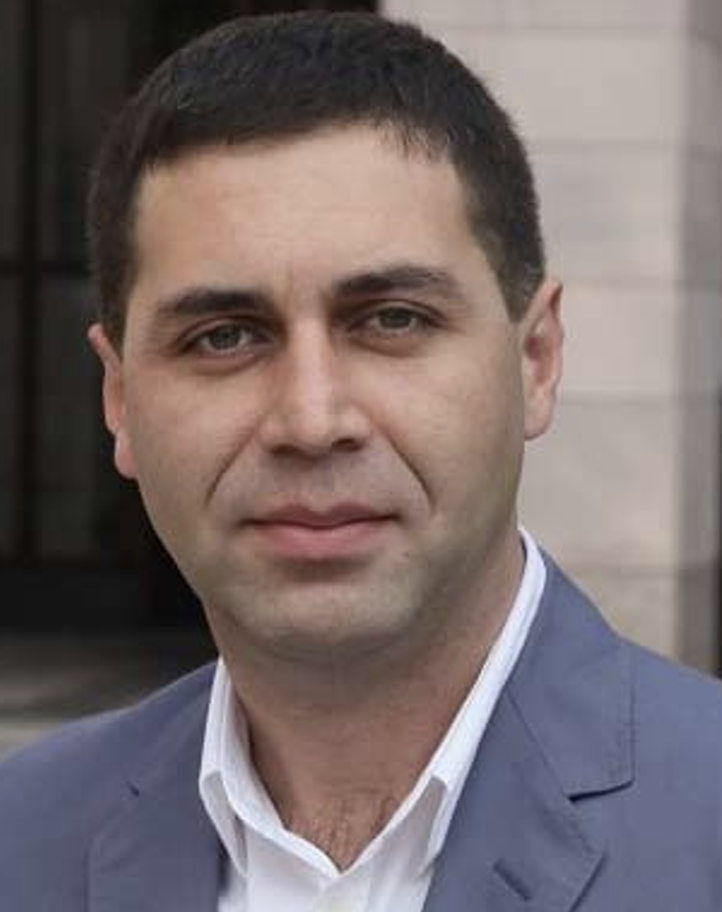}}]{Ozgur B. Akan (Fellow, IEEE)}
received the PhD from the School of Electrical and Computer En
gineering Georgia Institute of Technology Atlanta,
 in 2004. He is currently the Head of Internet of
 Everything (IoE) Group, with the Department of
 Engineering, University of Cambridge, UK and the
 Director of Centre for neXt-generation Communica
tions (CXC), Koc ¸ University, Turkey. His research
 interests include wireless, nano, and molecular com
munications and Internet of Everything.
\end{IEEEbiography}


\end{document}